\RequirePackage{fix-cm}
\documentclass[smallextended]{svjour3}       
\smartqed  
\usepackage{graphicx}

 \usepackage{mathptmx}      

\usepackage{amssymb,latexsym}

\usepackage{caption}
\usepackage[caption=false]{subfig}

\usepackage{latexsym}
\begin{document}

\title{The equatorial motion of the charged test particles in Kerr-Newman-Taub-NUT spacetime
}


\author{Hakan Cebeci   \and
      N\"{u}lifer \"{O}zdemir \and Se\c{c}il \c{S}entorun}


\institute{F. Author \at
             Department of Physics, Eski\c{s}ehir Technical University, 26470 Eski\c{s}ehir, Turkey\\
              \email{hcebeci@eskisehir.edu.tr , hcebeci@gmail.com}           
           \and
           S. Author \at
             Department of Mathematics,
Eski\c{s}ehir Technical University, 26470 Eski\c{s}ehir, Turkey \\
\email{nozdemir@eskisehir.edu.tr}
              \and
           T. Author \at
              Department of Physics, Eski\c{s}ehir Technical University, 26470 Eski\c{s}ehir, Turkey\\
              \email{secilo@eskisehir.edu.tr}
}

\date{Received: date / Accepted: date}

\maketitle

\begin{abstract}
In this work, we perform a detailed analysis of the equatorial motion of the charged test particles in Kerr-Newman-Taub-NUT spacetime. By working out the orbit equation in the radial direction, we examine possible orbit types. We investigate the conditions for existence of bound orbits in causality-preserving region as well as the conditions for existence of circular orbits for charged and uncharged particles. We also study the effect of NUT parameter on Newtonian orbits. Finally, we give exact analytical solutions of equations of equatorial motion for a charged test particle.
\end{abstract}

\section{Introduction}
\label{intro}

A remarkable solution to Einstein-Maxwell field equations is known as the Kerr-Newman-Taub-NUT (KNTN) spacetime which describes a rotating electrically charged source equipped with a gravitomagnetic monopole moment (also identified as the NUT charge) \cite{newman,demianski}. The spacetime includes four physical parameters; the gravitational mass, which is also called gravitoelectric charge; the gravitomagnetic mass (the NUT charge); the rotation parameter that is the angular speed per unit mass and the electric charge associated with the Maxwell field. In contrast to Kerr spacetime that is asymptotically flat, the rotating versions of spacetimes with gravitomagnetic monopole moment (Kerr-Taub-NUT and Kerr-Newman-Taub- NUT spacetimes) are asymptotically non-flat due to existence of the NUT charge. Although the Kerr-Taub-NUT and the KNTN spacetimes involve no curvature singularities, there exist conical singularities on the axis of symmetry. As discussed in \cite{misner}, one can get rid off conical singularities by imposing a periodicity condition over the time coordinate. However, this inevitably leads to the emergence of closed timelike curves in the spacetime that makes it unphysical in the context of causality. To make the spacetime with the NUT charge physically relevant, one can investigate the global analysis of such spacetimes as in \cite{bonnor,miller}. In this manner, an alternative physical interpretation of the spacetime with NUT charge has been given in \cite{bonnor} where the NUT metric is interpreted as a semi-infinite massless source of angular momentum (involving the singularity on the axis where $\theta=\pi$). Despite some unpleasing physical features of Taub-NUT spacetimes, the physical meaning of gravitomagnetic monopole moment and the physical interactions including NUT parameter have been comprehensively investigated. In \cite{bell1,bell2}, the physical meaning of NUT parameter has been exploited by examining the twisting effect of monopole moment on the orbit of the light rays. In \cite{bini1,bini2}, interaction of the massless scalar fields with gravitomagnetic monopole moment and gravitomagnetic effects regarding the NUT parameter have been investigated respectively. In \cite{aliev,esmer,liu}, some physical applications have been illustrated in the background of spacetimes involving the NUT parameter; namely in \cite{aliev}, gyromagnetic ratio related to KNTN spacetime has been obtained, in \cite{esmer}, hidden symmetries of Kerr-Taub-NUT spacetime in Kaluza-Klein theory have been explored, while in \cite{liu}, acceleration of particles on the background of Kerr-Taub-NUT spacetime has been studied. 

In order to detect the existence of a NUT source in the universe, one can either investigate the effect of this parameter on the motion of light or examine the effect on the motion of massive test particles. To accomplish and exploit such effect, one can study geodesics or the orbital motion on the background of spacetimes involving the NUT parameter. Such a study was initiated by \cite{zimmerman} where the Schwarzschild type geodesics on the background of NUT spacetime has been examined by concluding that such geodesics lie on the cones with the apex located at $r=0$. Later in \cite{kagramanova}, a comprehensive analytic investigation of complete and incomplete geodesics in a (non-rotating) Taub-NUT spacetime has been realised. On the other hand, the motion of particles in a Kerr-Taub-NUT gravitational source immersed in a magnetic field has been investigated in \cite{abdujabbarov1}. In addition, in \cite{abdujabbarov2}, energy extraction process (Penrose process) has been examined in rotating Kerr-Taub-NUT spacetime. In the work \cite{grenzebach}, an analytic expression for shadows (known as a special lensing property) of a Kerr-Newman-NUT spacetime has been obtained while examining the motion of photon in this background. In a recent work \cite{pradhan}, circular geodesics of uncharged test particles has been analysed in KNTN spacetime while in \cite{jefremov}, timelike circular geodesics in (non-rotating) NUT spacetime has been studied while also discussing the Von Zeipel cylinders with respect to stationary observers and determining the relation of such cylinders to inertial forces. Recently the motion of charged particles has been investigated in an Einstein-Maxwell spacetime with NUT parameter \cite{clement}, where in such a spacetime the causality violation has also been examined. Very recently in \cite{mukherjee}, the equations of motion of a test particle have been examined in a special class of Kerr-Newman-NUT spacetime background where a specific relation is imposed on the NUT charge, electric charge and rotation parameter of the spacetime. In addition to study of geodesics of light and massive test particles in spacetimes with NUT parameter, gravitational waves can also be viewed as a possible third method to detect a NUT charge in the universe \cite{abbott}.

This work is devoted to the study of equatorial orbits of charged massive test particle in KNTN spacetime. In previous works, the motion of charged massive test particles has been analytically investigated in Reissner-Nordstr\"{o}m \cite{grunau,pugliese1}, Reissner-Nordstr\"{o}m-(Anti)-de-Sitter \cite{olivares}, Kerr-Newman \cite{hackmann1} and Kerr-Newman-AdS (in $f(R)$ gravity) \cite{soroushfar} respectively. Our aim is to make a detailed analysis of the equatorial motion of the charged massive test particles in KNTN spacetime and to investigate the effect of the rotation and the NUT parameters on the equatorial motion. In fact, we have recently examined the non-equatorial orbital motion of charged massive test particles in KNTN background in \cite{cebeci1}, where we have briefly mentioned the conditions for the existence of the equatorial orbits in rotating NUT spacetime without making a detailed analysis and investigation of such orbits. In this work, we present a more elaborate analysis of the equatorial motion of charged massive test particles in KNTN spacetime where we particularly examine the existence of bound and circular orbits over equatorial plane. We should remark that, the study of equatorial orbits in rotating NUT spacetime deserves a separate care and investigation. As is also mentioned in \cite{bini2}, the equatorial orbits in rotating NUT spacetimes do not exist for arbitrary rotation and the NUT parameters. In \cite{cebeci1}, we have shown that, there exist equatorial orbits in such spacetimes provided that either the NUT parameter should vanish or a certain relation between the angular momentum and the energy of the test particle and the rotation parameter should exist (for arbitrary NUT parameter). In this work, we concentrate on the latter, i.e. we examine the equatorial orbits of the charged test particles in which such a relation holds. We examine the possible orbit types depending on the value of the energy of the test particle while a special investigation is devoted to the study of the existence of bound orbits in causality-preserving region and existence of circular orbits. To our knowledge, the study of circular orbits in rotating spacetimes has been initiated by Bardeen et al. \cite{bardeen} for Kerr spacetime. Later, the investigation of the existence of circular geodesics has been accomplished in Kerr-Newman spacetime \cite{dadhich,pugliese2}. A detailed investigation of such orbits has also been realised in Kerr-de Sitter and Kerr-Anti-de Sitter spacetimes including cosmological constant together with mass and rotation parameters \cite{stuchlik_1,stuchlik_2}. In our study, we derive necessary conditions for the existence of equatorial bound orbits in causality-preserving region as well as the existence circular orbits in KNTN spacetime. In addition, we study the effect of the NUT parameter on the equatorial Newtonian orbits. Finally, we present the analytical solutions of the equations of motion over the equatorial plane by expressing them in terms of Weierstrass $\wp$, $\sigma$, and $\zeta$ functions. We also provide plots of possible orbit types and calculate the perihelion shift for an equatorial bound orbit.

Organisation of the paper is as follows: In Section 2, we provide an introduction to KNTN spacetime. In Section 3, we obtain the governing equations of equatorial motion of the charged test particles. In Section 4, we make a comprehensive analysis of possible orbit types. In the same section, we examine the conditions for the existence of equatorial bound orbits in causality-preserving region and existence of circular orbits. Moreover, the Newtonian limit of the equatorial orbits are discussed while investigating the physical effect of the NUT parameter on the Newtonian orbits as well. In Section 5, we present the exact analytic solutions of the equatorial orbits while also calculating the perihelion shift for an equatorial bound orbit. We end up with some comments and conclusions.

\section{Kerr-Newman-Taub-NUT spacetime}

The KNTN spacetime is known as a stationary rotating solution of the Einstein-Maxwell field equations. The metric is asymptotically non-flat due to the existence of a NUT charge which is also identified as the gravitomagnetic monopole moment. In Boyer-Lindquist coordinates, KNTN spacetime can be written as (with asymptotically non-flat structure),
\begin{equation}
g = - \frac{\Delta}{\Sigma} (dt- \chi d \varphi )^{2} + \Sigma
\left(\frac{d r^{2} }{\Delta} + d \theta ^{2} \right) + \frac{\sin
^{2} \theta }{\Sigma } \left( a d t - ( r^{2} + \ell^{2} + a^{2} )
d \varphi \right)^{2}
\label{Kerr_1}
\end{equation}
where
\begin{eqnarray}
\Sigma &=& r^{2} + (\ell + a \cos \theta)^{2} , \nonumber \\
\Delta&=& r^{2} - 2 M r + a^{2} - \ell^{2}+Q^2 , \label{Kerr_2}\\
\chi &=& a \sin^{2} \theta - 2 \ell \cos \theta .\nonumber
\end{eqnarray}
Here, $M$ can be identified as the parameter related to the physical mass of the gravitational source, $a$ is associated with its angular momentum per unit mass while $\ell$ specifies gravitomagnetic monopole moment
of the source which is also identified as the NUT charge. Also, $Q$ is specified as the electric charge. The electromagnetic field of the source can be expressed as the potential 1-form
\begin{equation}
A=A_\mu dx^\mu=- \frac{Qr}{\Sigma}(dt- \chi d \varphi).\label{Kerr_3}
\end{equation}
We use geometrized units such that $c=1$ and $G=1$. As is also remarked in \cite{miller}, there exist two types of singularities related to that spacetime; one coming from the singularities of the metric components and the other resulting from vanishing of the determinant of the metric. The former results in the singularities at the spacetime coordinates where $\Delta(r)=0$ and $\Sigma(r,\theta)=0$ producing singularities at the horizons $r=r_{\mp}$ and a conical singularity (at $r=0$ and $\cos \theta=-\frac{\ell}{a}$ with $\ell^2<a^2$) respectively. The latter occurs at $\theta=0$ and $\theta=\pi$. When the NUT parameter $\ell=0$, the conical singularity obviously turns into the equatorial ring singularity at $r=0$ and $\theta=\frac{\pi}{2}$. We further remark that, the metric singularities related to radial coordinate exist at the locations
\begin{equation}
r_{\pm}=M\pm\sqrt{M^2-a^2+\ell^2-Q^2} ,\label{Kerr_4}
\end{equation}
where $\Delta(r_\pm)=0$ provided that $M^2-a^2+\ell^2-Q^2 \geq 0$. The singularity $r_{-}$ can be identified as inner (or Cauchy) horizon while $r_{+}$ can be named as outer (or event) horizon. It can also be seen that, the spacetime allows a family of locally non-rotating observers which rotate with coordinate angular velocity given by
\begin{equation}
\Omega=-\frac{g_{t\varphi}}{g_{\varphi\varphi}}=\frac{\Delta \chi - a \sin^2 \theta (r^{2} + \ell^{2} + a^{2})}{\Delta \chi^2 - \sin^2 \theta (r^{2} + \ell^{2} + a^{2})^2}\label{Kerr_5}
\end{equation}
which is known as the frame dragging effect arising due to the presence of the off-diagonal component $g_{t \varphi}$ of the metric. We also remark that at the outermost singularity $r_{+}$, the angular velocity can be calculated as
\begin{equation}
\Omega_{+}=-\left.\left(\frac{g_{t\varphi}}{g_{\varphi\varphi}}\right)\right|_{r=r_{+}}=\frac{a}{r_{+}^{2} + \ell^{2} + a^{2}} \, \cdot \label{Kerr_6}
\end{equation}
It can also be easily seen that the Killing vectors $\xi_{(t)}$ and $\xi_{(\varphi)}$ generate two constants of motion namely the energy and the angular momentum of the test particle. Moreover, it is straightforward to show that the Killing vector $\xi=\xi_{(t)}+\Omega_{+}\xi_{(\varphi)}$ becomes null at the metric singularity where $r=r_{+}$.

\section{The equations of motion}

In this section, we examine the motion of charged test particle over the equatorial plane in KNTN spacetime. Traditionally, to obtain the equation of motions over the equatorial plane, one usually starts with the Lagrangian expression \cite{chandrasekhar} and simply substitute $\theta=\frac{\pi}{2}$ in the resulting expressions. However, this standard technique cannot be directly applied for the KNTN spacetime since as is explicitly illustrated in \cite{bini2} and \cite{cebeci1}, the existence of the equatorial orbits requires either the vanishing of the NUT parameter ($\ell=0$) or a specific relation between the rotation parameter $a$, the energy $\bar{E}$, the angular momentum $\bar{L}$ of the test particle to hold. For that reason, one cannot directly substitute $\theta=\frac{\pi}{2}$ for the equatorial orbital motion. Instead, to get the field equations over the equatorial plane, one should start with the celebrated Hamilton-Jacobi equation, and then obtain the condition for the existence of equatorial orbits and substitute such a relation in the remaining governing equations. Therefore, governing orbit equations over equatorial plane can be obtained by using Hamilton-Jacobi method discussed in \cite{cebeci1}. To summarize, one can start with Hamilton-Jacobi equation \cite{carter_1,carter_2,kamran,frolov}
\begin{equation}
2 \frac{\partial S}{\partial \tau}=g^{\mu \nu} \left( \frac{\partial S}{\partial x^{\mu}}-q A_{\mu}\right) \left( \frac{\partial S}{\partial x^{\nu}}-q A_{\nu}\right) \label{motion_1}
\end{equation}
where $\tau$ denotes an affine parameter and $q$ is the electric charge of the test particle. Also, the existence of the timelike Killing vector $\xi_{(t)}$ and spacelike Killing vector $\xi_{(\varphi)}$ for the KNTN spacetime (\ref{Kerr_1})  lead to the identifications 
\begin{equation}
P_{t}=-E, \qquad P_{\varphi}=L, \label{motion_6}
\end{equation}
where the expression for the canonical momenta $P_{\mu}$ can be written as
\begin{equation}
P_{\mu}=\frac{\partial S}{\partial x^{\mu}}=m g_{\mu \nu} \frac{d x^{\nu}}{d \tau}+q A_{\mu} \label{motion_5}.
\end{equation}
Here $m$ is the mass of the test particle while $E$ and $L$ correspond to the energy and the angular momentum of the test particle. Now, it has been shown in \cite{cebeci1} that if the relation 
\begin{equation}
\bar{L}=\frac{a(2 \bar{E}^2-1)}{2 \bar{E}} \, ,  \qquad (\bar{E}\neq0), \label{motion_22}
\end{equation}
with
\begin{equation}
\bar{E}=\frac{E}{m} \, , \qquad  \bar{L}=\frac{L}{m} 
\end{equation}
is imposed between the rotation parameter $a$, rescaled energy $\bar{E}$ and the angular momentum $\bar{L}$ of the test particle, the particle is confined to move over the equatorial plane. One can infer from this relation that if $ a \neq 0$, but $2 \bar{E}^2=1$ ($\bar{L}=0$), the orbits of the charged test particle can be identified as the motion with vanishing orbital angular momentum. One more crucial remark is that, if $2 \bar{E}^2>1$, $\bar{L}>0$ implying that equatorial orbits are direct (or prograde) orbits ($\bar{L}$ and $a$ have the same sign also assuming that $a>0$). If on the other hand, $2 \bar{E}^2<1$, $\bar{L}<0$ which implies that the orbits are retrograde orbits ($\bar{L}$ and $a$ have opposite signs). 

\noindent In addition, to get the complete governing equations of motion over the equatorial plane, one can introduce a new time parameter $\lambda$ as in \cite{zakharov,mino} such that
\begin{equation}
\frac{d\lambda}{d\tau}=\frac{1}{\Sigma} \label{motion_14}
\end{equation}
and substitute $\theta= \frac{\pi}{2}$, $ \bar{L}= \frac{a(2 \bar{E}^2-1)}{2 \bar{E}}$ together with Carter constant $\frac{K}{m^2}=\ell^2+\frac{a^2}{4 \bar{E}^2}$
in the orbit equations presented in \cite{cebeci1}. To conclude, the governing orbit equations take the following form:
\begin{equation}
\left(\frac{dr}{d\lambda}\right)^2=\bar{P}_r(r),\label{motion_24}
\end{equation}
\begin{equation}
\left( \frac{dr}{d \varphi} \right)^2 = \frac{\Delta^2 (r)\bar{P}_r(r)}
{a^2\left(\left[\bar{E} (r^2  + \ell^2)+\frac{a^2}{2\bar{E}}-\bar{q}Q r\right]-\frac{\Delta(r)}{2\bar{E}}\right)^2},\label{motion_25}
\end{equation}
\begin{equation}
\left(\frac{dr}{dt}\right)^2=\frac{\Delta^2 (r)\bar{P}_r(r)}{\left((r^2 +a^2 + \ell^2) \left[\bar{E} (r^2  + \ell^2)+\frac{a^2}{2\bar{E}}-\bar{q}Q r\right]-\frac{a^2\Delta(r)}{2\bar{E}}\right)^2}\label{motion_26}
\end{equation}
where we define
\begin{equation}
\bar{P}_r(r)=\left[\bar{E}(r^2+\ell^2)+\frac{a^2}{2 \bar{E}}-\bar{q} Q r\right]^2-\Delta(r) \left(r^2+\ell^2+\frac{a^2}{4\bar{E}^2}\right). \label{motion_27}
\end{equation}

\noindent Here $ \bar{q} = \frac{ q }{m} $. Now writing the radial potential $\bar{P}_r(r)$  over the equatorial plane in the form 
\begin{equation}
\bar{P}_r(r)=r^4 \left(\bar{E} \left( 1+ \frac{\ell^2}{r^2}\right)+ \frac{a^2}{2 \bar{E} r^2} -\frac{\bar{q}Q}{r}\right)^2 -\Delta(r) \left(r^2+\ell^2+\frac{a^2}{4\bar{E}^2}\right) ,
\label{motion_27_a}
\end{equation}
one can physically interpret the term $\frac{a^2}{2 \bar{E} r^2}$ as the spin orbit coupling potential arising from the orbital motion of the test particle around the rotating spacetime (due to relation $\frac{a \bar{L}}{r^2} = \frac{a^{2}}{r^{2}} (\bar{E} - \frac{ 1 }{2 \bar{E} } ) $ over equatorial plane) and term $\frac{\bar{q} Q}{r}$ as electrostatic interaction potential (between charge of test particle and charge of the spacetime). 

\noindent Finally, we note that the equations (\ref{motion_24})-(\ref{motion_26}) have been obtained under the assumption that $ \ell \neq 0 $. When $\ell=0$ (vanishing of the NUT parameter), one does not require the relation (\ref{motion_22}), 
since relation (\ref{motion_22}) should be used for the existence of equatorial orbits in a spacetime with NUT parameter (i.e in a spacetime with $\ell \neq 0$). Therefore, for $ \ell = 0 $, one should consider the equations of motion outlined in \cite{cebeci1}. Indeed, if one substitutes $ \ell = 0 $ (taking $\theta=\frac{\pi}{2}$ for the equatorial orbits) in the equations of motion presented in \cite{cebeci1}, one obtains the orbit equations for the charged test particle in Kerr-Newman spacetime where in this case, the angular momentum $ \bar{L} $ and energy $ \bar{E} $ of the test particle appear as independent parameters of the radial potential (see also \cite{hackmann1}).

\subsection{Ergoregion in KNTN spacetime}

\noindent Before moving to next section, we should note that there exists another feature of KNTN spacetime that is worth mentioning. Such a property that deserves special investigation is the existence of an ergoregion (or existence of ergosurface) and it is seen as characteristic of stationary spacetimes. The existence of an ergoregion in stationary spacetimes require that norm of timelike Killing field (i.e metric component $g_{tt}$) becomes positive. It means that the coordinate $ t $ in this region is no longer a timelike coordinate but it becomes spacelike. We note that in recent works \cite{pugliese_1} and \cite{pugliese_2}, the characteristics of such a region on the equatorial plane of Kerr spacetime has been examined in detail. For KNTN spacetime, in the region where the relation
\begin{equation}
\Delta- a^2 \sin^{ 2 } \theta <0
\end{equation}
holds, the metric component $ g_{tt}$ becomes positive.  
Therefore, the location of ergosurface for KNTN spacetime, also known as the stationary limit surface, can be obtained from 
\begin{equation}
r^{2} - 2 M r + a^{2} \cos^{ 2 } \theta + Q^{2}  - \ell^{2} = 0   
\end{equation}
whose solution leads to
$$
r_{e}^{\mp} (\theta) = M \mp \sqrt{ M^{2} + \ell^{2} - Q^{2} - a^{2} \cos^{2} \theta } .
$$
Here $ r_{e}^{-} ( \theta ) $ can be identified as inner ergosurface while $ r_{e}^{+} ( \theta ) $ is identified as outer ergosurface. 
One can immediately see that the relation
\begin{equation}
r_{e}^{-} < r_{-} < r_{+} < r_{e}^{+} 
\end{equation}
holds between inner and outer ergosurfaces and Cauchy (inner horizon $ r_{-} $ ) and event (outer horizon $ r_{+} $) horizons. We note that ergosurfaces coincide with Cauchy and event horizons when $ \theta = 0 $ and $ \theta = \pi $. When $ \theta = \frac{\pi}{2} $, i.e over the equatorial plane of KNTN spacetime, expression for ergosurfaces takes the form 
\begin{equation}
r_{e}^{\mp} = M \mp \sqrt{ M^{2} + \ell^{2} - Q^{2}} \, .
\end{equation}
It is interesting to observe that ergosurfaces over equatorial plane of KNTN spacetime are independent of spacetime rotation parameter $ a $ but they depend on other spacetime parameters. 
\noindent Another remarkable feature of the ergoregion is that static observers cannot exist in ergoregion (where 
$ r_+ < r < r_{e}^{+}  $) of equatorial KNTN spacetime.  It means that there is no static observer with $ \frac{d r}{d \tau} = 0 $, $ \frac{ d \theta }{d \tau} = 0 $ and $ \frac{ d \varphi}{d \tau} = 0 $ where $ \tau $ denotes proper time. Referring to our previous work \cite{cebeci1} , this fact can be seen from angular velocity expression 
\begin{equation}
\frac{d \varphi}{ d \tau} = \frac{ a }{(r^{2} + \ell^{2} ) \Delta } \left( \frac{1}{ 2 \bar{E} } ( a^{2} - \Delta ) +  \bar{E} ( r^{2} + \ell^{2} ) - \bar{q} Q r  \right) 
\end{equation}
such that expression never vanishes in ergoregion where $ r_{+} < r < r_{e}^{+} $. Looking at this expression, owing to  
$$
a^{2} - \Delta > 0
$$
in the ergoregion $ r_{+} < r < r_{e}^{+} $ over equatorial plane, expression is strictly positive for an uncharged particle ($\bar{q} = 0$) that enters into ergoregion with positive energy. For a charged particle with positive energy, expression is positive provided that
\begin{equation}
4 \bar{E}^{2} \ell^{2} - \bar{q}^{2} Q^{2} > 0 \, .
\end{equation}
In addition, angular velocity expression has same sign with spacetime rotation parameter $ a $ which implies that test particles that enter into ergoregion are forced to rotate in direction of rotation of KNTN black hole. Obviously, it is zero when $ a = 0 $ where spacetime is static in that case.  

\noindent Also, another characteristics of ergoegion that deserves further mentioning is that Killing vector $\xi_{(t)}=\frac{\partial}{\partial t}$ becomes spacelike \cite{wald,chandrasekhar}. Physically, it means that, in the ergoregion, energy $\bar{E}$ of the test particle measured with respect to an observer at spatial infinity can be negative. Then from physical point of view, an immediate consequence of existence of such a region between event horizon and stationary limit surface is that it can allow Penrose process that results in the extraction of energy from KNTN black hole \cite{abdujabbarov2}. In addition, it should be noted that negative energy particles within ergoregion cannot exit from ergoregion, while particles with positive energy can enter that region and exit from ergoregion. The fact that energy of test particle in ergoregion can be negative also requires that angular momentum $ \bar{L} $ of particle measured with respect to an observer at spatial infinity can be negative as well. Especially, if one considers circular motion in ergoregion, the energy $ \bar{E} $ and angular momentum $ \bar{L} $ turns out to be negative \cite{chandrasekhar,pugliese_1} .
As a final comment, for the motion over equatorial plane of KNTN spacetime, the negativity of angular momentum also agrees with constraint relation (\ref{motion_22}) such that when $ \bar{E} < 0 $ (provided that $ 2 \bar{E}^{2} > 1 $), it requires that $ \bar{L} < 0 $ as well. At this point, we should point out that a more detailed investigation of motion in ergoregion of KNTN spacetime as done for Kerr spacetime in \cite{pugliese_1,pugliese_2}, could be subject of another future work.

\section{Analysis of the orbit configurations}

In this section, we make a classification of possible orbit configurations with respect to radial motion expressed by the radial potential $\bar{P}_r(r)$.  Next, we make a detailed investigation of the existence of bound orbits in causality-preserving region and existence of circular orbits. Finally, we illustrate the effect of the NUT parameter on the equatorial Newtonian orbits.

\noindent First, one can easily see that radial potential $\bar{P}_r(r)$ is a fourth order polynomial in $r$ with real coefficients, where it can be expressed in the form
\begin{equation}
\bar{P}_r(r)=B_0+B_1 r+B_2 r^2+B_3 r^3+B_4 r^4 \label{radial_potential_1}
\end{equation}
where the coefficients read
\begin{equation}
B_4= \bar{E}^2-1, \label{radial_potential_2}
\end{equation}
\begin{equation}
B_3=2(M-\bar{E} \bar{q}Q), \label{radial_potential_3}
\end{equation}
\begin{equation}
B_2=2 \bar{E}^2\ell^2+Q^2(\bar{q}^2-1)-\frac{a^2}{4 \bar{E}^2}, \label{radial_potential_4}
\end{equation}
\begin{equation}
B_1= 2 (M-\bar{q} Q \bar{E})  \left(  \ell^2+\frac{a^2}{4 \bar{E}^2} \right)-  \frac{\bar{q} Q a^2}{2 \bar{E}}, \label{radial_potential_5}
\end{equation}
and
\begin{equation}
B_0=\ell^4 \bar{E}^2+(\ell^2-Q^2)\left( \ell^2+\frac{a^2}{4 \bar{E}^2} \right). \label{radial_potential_6}
\end{equation}
Also we note that the radial motion is possible if $\bar{P}_r(r) \geq 0$. Then, according to the roots of the radial polynomial $\bar{P}_r(r)$, one can identify the following orbit types in general \cite{kagramanova,oneil}:\\

\noindent {\bf i.} Bound orbit: When the particle moves in a region $r_2 \leq r \leq r_1$ (either $r_1>r_2>0$ or $r_2<r_1<0$), the motion of the particle can be identified as bound where the point $r=0$ is not crossed. On the other hand if the bound orbit exists in a region $r_2<r<r_1$ with $r_2<0$ and $r_1>0$, the orbit can be identified as crossover bound orbit where the point $r=0$ is crossed twice. In addition, the bound orbit can be identified as many-world bound orbit \cite{grunau,grunau2}, if $r_2<r_- $ and $r_1>r_+$. In such a case the test particle moves in a bound orbit stretching from one part of the spacetime region into another part several times where particle crosses two metric singularities at $r=r_-$ and $r=r_+$ and turns back at $r=r_2$. 
 The bound orbits are possible if $\bar{P}_r(r)$ has four real roots or two real roots (with two complex roots) or two real double roots or one real triple root and one real root. In such cases, depending on the real roots of the radial potential, there may exist one or two bound regions. \\
  
\noindent {\bf ii.} Circular Orbit: The orbit is called circular if $\bar{P}_r(r)$ has a real double root at $r=r_c$ where $r_c$ denotes the radius of the circular orbit.\\

\noindent {\bf iii.} Escape orbit: The orbit is called escape if the particle moves either in the range $r_1<r<\infty$ (with $r_1>0$) or $-\infty<r<r_2$ (with $r_2<0$) where the point $r=0$ is not crossed. The orbit can be identified as crossover escape orbit if the particle moves either in the range $r_1<r<\infty$ (with $r_1<0$) or $- \infty <r <r_2$ (with $r_2>0$). In such a case the particle crosses the point $r=0$ twice. If the motion of the particle is restricted to be either in the interval $r_1<r<\infty$ with $r_1<r_-<r_+$ or $-\infty<r<r_2$ with $r_-<r_+<r_2$, then the orbit can be referred as two-world escape orbit. In such a case, the particle passes through from one part of the spacetime region into another part where again the particle will cross two metric singularities twice. Likewise, escape orbits can arise when $\bar{P}_r(r)$ has four real roots or two real roots (with two complex roots) or two real double roots or one real triple root and one real root. We note that there may exist one or two escape orbits depending on the number of real roots.\\

\noindent {\bf iv.} Transit orbit: The orbit is said to be transit if the particle starts from $\mp \infty$, crosses $r=0$ and moves to $\pm \infty$. This can be possible if $\bar{P}_r(r)$ has no real roots. It is obvious that if $\bar{P}_r (0)>0 $, the particle can cross the point  $r=0$. This can happen if 
\begin{equation}
B_0=\ell^4 \bar{E}^2+(\ell^2-Q^2)\left( \ell^2+\frac{a^2}{4 \bar{E}^2} \right) >0 . 
\end{equation}
It is seen that if $\ell \geq Q$, the particle can always cross the point $r=0$.\\

\noindent  
 One can further examine the possible orbit configurations depending on the value of the energy of the test particle:\\

\noindent{\bf 1.} The case for $\bar{E}^2 \neq 1$:\\

\noindent If $\bar{P}_r (r)$ has four different real roots, one can obtain two bound orbits for $\bar{E}^2<1$, while for $\bar{E}^2>1$ one can get one bound and two escape orbits. If $\bar{P}_r (r)$ has two different real zeros (and two complex conjugate roots), then one can get only one bound orbit for $\bar{E}^2<1$, while for $\bar{E}^2>1$, one can obtain two escape orbits. If $\bar{P}_r (r)$ has no real zeros, then the radial motion is not possible for $\bar{E}^2<1$ since $\bar{P}_r (r)<0$ for all $r$, while one can get a transit orbit for $\bar{E}^2>1$ since $r\rightarrow \mp \infty$, $\bar{P}_r(r)\rightarrow \infty$ . 
\\

\noindent{\bf 2.} The case for $\bar{E}^2 = 1$:\\

\noindent For $\bar{E}^2=1$, $\bar{P}_r(r)$ becomes a third order polynomial. Moreover, the orbit configurations can change according to the sign of the coefficient of the first term (i.e. the coefficient of $r^3$). For both of the cases $M>\bar{q}Q$ and $M<\bar{q}Q$, if $\bar{P}_r(r)$ has three distinct real roots, there exist one bound and one escape orbits. If on the other hand $\bar{P}_r(r)$ has only one real root (together with two complex roots), then there exists only one escape orbit. For this case, for $M>\bar{q}Q$, escape orbit is observed in the interval $r_1\leq r < + \infty$ while for $M<\bar{q}Q$, the escape orbit can be realised in the interval $-\infty< r \leq r_1$ where $r_1$ is the real root of $\bar{P}_r(r)$. 

\noindent It is also of interest to examine the case where $\bar{E}^2=1$ and $M=\bar{q}Q$. For this special case, $\bar{P}_r(r)$ becomes a second order polynomial:
\begin{equation}
\left( 2\ell^2+Q^2 (\bar{q}^2-1)-\frac{a^2}{4}\right) r^2-\frac{a^2}{2} \bar{q} Q r+\ell^2+(\ell^2-Q^2)\left(\ell^2+\frac{a^2}{4}\right)=0. \label{analysis_19}
\end{equation}
In this case, the orbit configurations can modify according to the sign of the coefficient of $r^2$ term. Then, if
\begin{equation}
2 \ell^2+Q^2 (\bar{q}^2-1)-\frac{a^2}{4}>0, \label{analysis_20}
\end{equation}
one can obtain two escape or a transit orbit according to whether $\bar{P}_r(r)$ has two different real zeros (or one double zero) or no zeros respectively. If on the other hand,
\begin{equation}
2 \ell^2+Q^2 (\bar{q}^2-1)-\frac{a^2}{4}<0, \label{analysis_21}
\end{equation}
one can get one bound orbit or no radial motion according to whether $\bar{P}_r(r)$ has two different real zeros or no zeros respectively.

\begin{figure*}[htp]
\centering
\subfloat[Effective potential for radial motion of positively charged test particle ($\bar{q}=10$)]{\label{nut_parameter_energy_greater_than_one_positive_charge}
\includegraphics[width=0.6\linewidth]{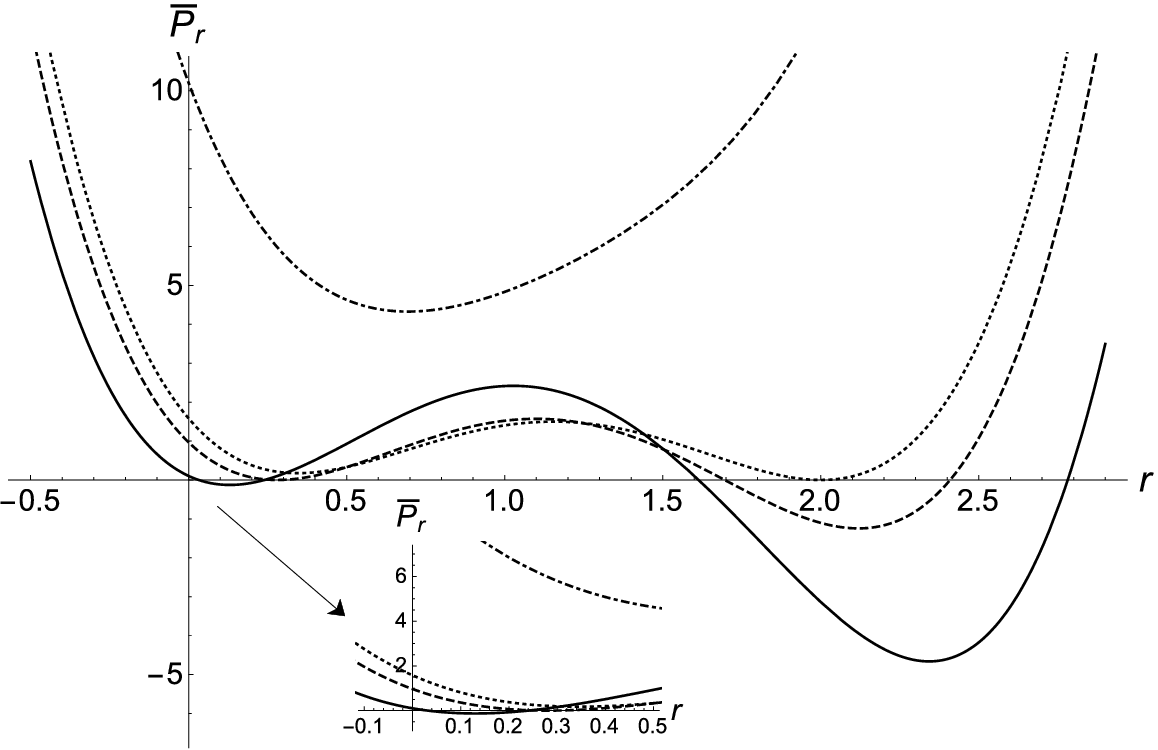}}\\
\subfloat[Effective potential for radial motion of negatively charged test particle ($\bar{q}=-5$) ] {\label{nut_parameter_energy_greater_than_one_negative_charge} 
\includegraphics[width=0.45\linewidth]{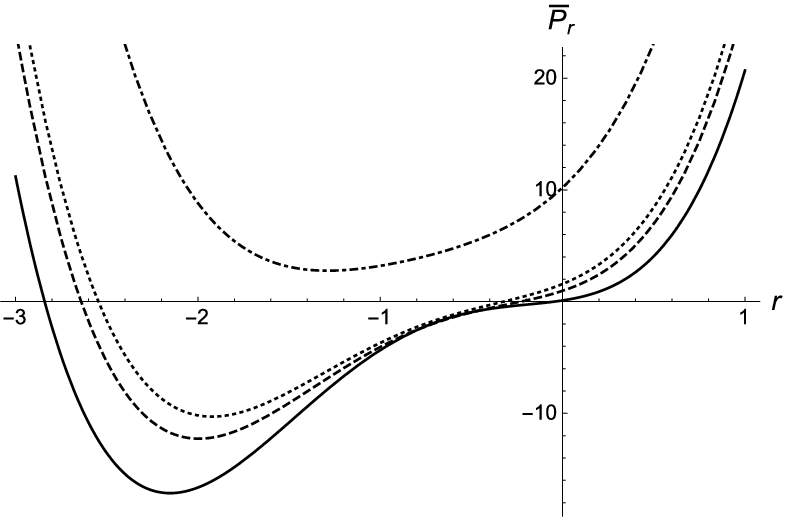}  } \hspace{0.5cm}
\subfloat[Effective potential for radial motion of  uncharged test particle ($\bar{q}=0$)]{ \label{nut_parameter_energy_greater_than_one_zero_charge}
 \includegraphics[width=0.45\linewidth]{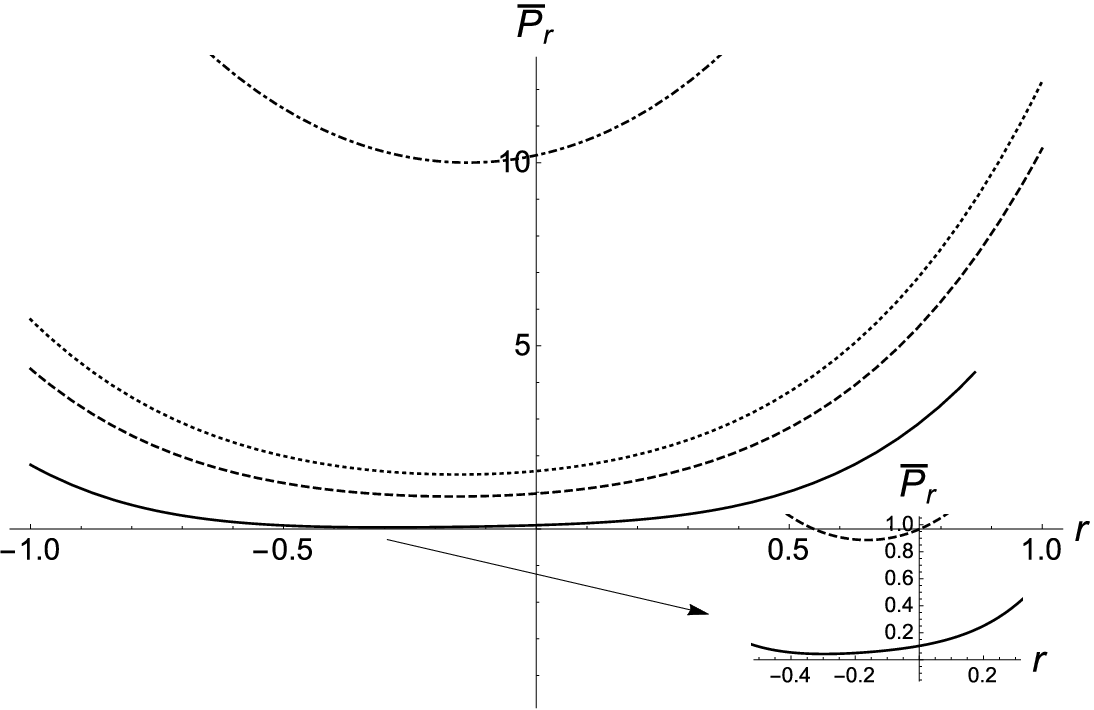}  }
\caption{Effective radial potentials for different values of NUT parameter $\ell$ when $\bar{E}>1$. Black solid, black dashed, black dotted and black dot-dashed curves are obtained for $\ell=0.4$, $\ell=0.672720$, $\ell=0.758142$ and $\ell=1.2$ respectively. In all the plots, we take  $\bar{E}=2$, $M=1$, $Q=0.4$ and $a=0.9$.}\label{orbit_configurations_1}
\end{figure*}

\begin{figure*}[htp]
    \centering   
    \subfloat[Effective potential for radial motion of positively charged test particle ($\bar{q}=5$)]
 {\label{nut_parameter_energy_less_than_one_positive_charge}
\includegraphics[width=0.42\linewidth]{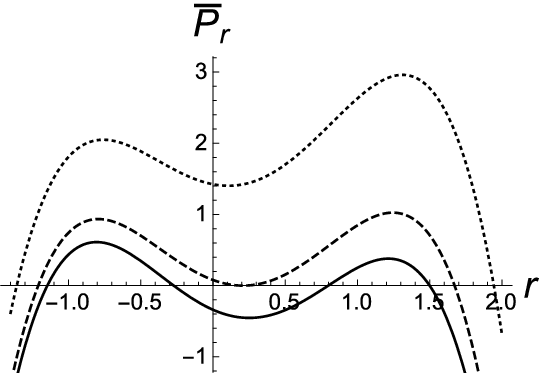}  }  \qquad
   \subfloat[Effective potential for radial motion of negatively charged test particle ($\bar{q}=-2$)] 
   {\label{nut_parameter_energy_less_than_one_negative_charge}
\includegraphics[width=0.42\linewidth]{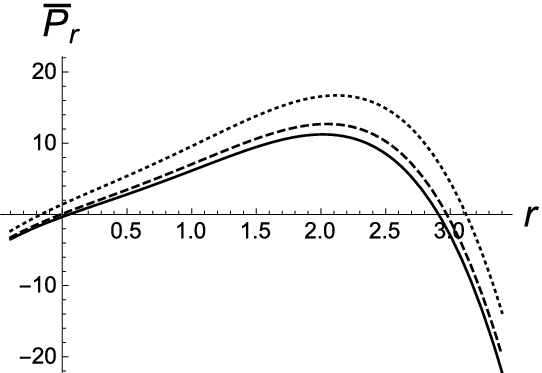}  }\\
   \subfloat[Effective potential for radial motion of uncharged test particle ($\bar{q}=0$)]
{\label{nut_parameter_energy_less_than_one_zero_charge}
\includegraphics[width=0.5\linewidth]{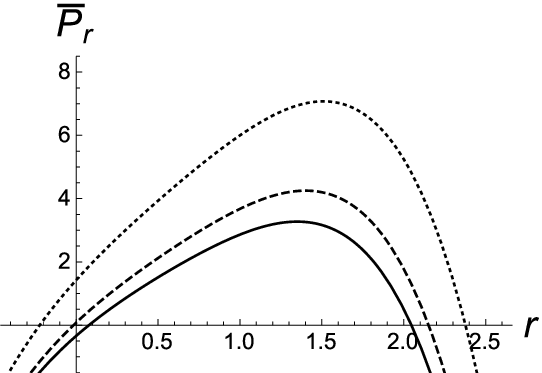}  }   
          \caption{Effective radial potentials for different values of NUT parameter $\ell$ when $\bar{E}<1$. Black solid, black dashed and black dotted curves are obtained for $\ell=0.1$, $\ell=0.437329$ and $\ell=0.8$ respectively. In all the plots, we take  $\bar{E}=0.3$, $M=1$, $Q=0.4$ and $a=0.9$.}
\label{orbit_configurations_2}
\end{figure*}

\begin{figure*}[htp]
    \centering
     \subfloat[Effective potential for radial motion of positively charged test particle ($\bar{q}=10$)]
  {\label{nut_parameter_energy_equal_to_one_positive_charge}
\includegraphics[width=0.65\linewidth]{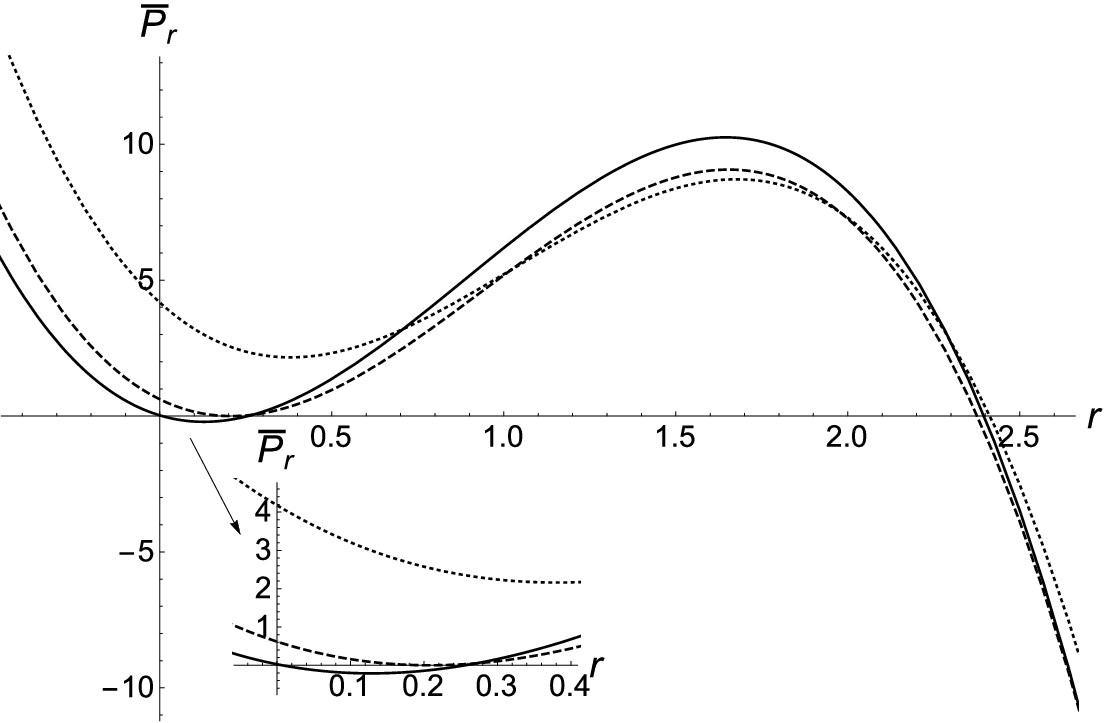}  }  \\  
 \subfloat[Effective potential for radial motion of negatively charged test particle ($\bar{q}=-12$)]
 {\label{nut_parameter_energy_equal_to_one_negative_charge}
\includegraphics[width=0.42\linewidth]{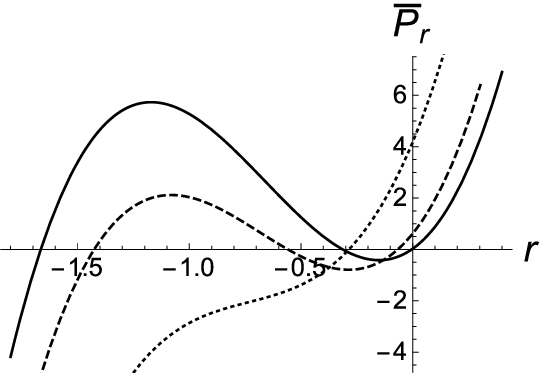}  }  \qquad
\qquad \subfloat[Effective potential for radial motion of uncharged test particle ($\bar{q}=0$)]
  {\label{nut_parameter_energy_equal_to_one_zero_charge}
\includegraphics[width=0.42\linewidth]{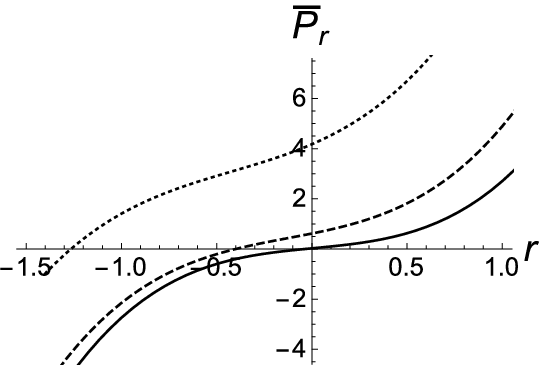}  } 
 \caption{Effective radial potentials for different values of NUT parameter $\ell$ when $\bar{E}=1$. Black solid, black dashed and black dotted curves are obtained for $\ell=0.4$, $\ell=0.747708$ and $\ell=1.2$ respectively. In all the plots, we take $M=1$, $Q=0.4$ and $a=0.9$.}\label{orbit_configurations_3}
\end{figure*}

\noindent As a final remark, to see the effect of NUT parameter on the formation of possible orbit configurations, it will be useful to obtain the plots of the effective potential $\bar{P}_r(r)$ for different values of NUT parameter and to make an analysis of possible orbit configurations as NUT parameter changes. Such plots are given in Figures \ref{orbit_configurations_1}-\ref{orbit_configurations_3}.
In these plots, the parameters can be chosen to obtain a physically acceptable radial motion (i.e $\bar{P}_r(r) \geq0$). Looking at plots, one can observe that for case $ \bar{E}^{2} > 1 $, for positively charged particle, while bound and circular orbits are formed for sufficiently small and critical values of NUT parameter (for the critical values $\ell=0.67272$ and $\ell= 0.758142$ of the NUT parameter circular orbits can form), transit orbits are observed as $ \ell $ increases. However, for negatively charged particle, (for same energy and spacetime parameters chosen) while escape orbits are formed for small values of NUT parameter, as in motion of charged test particle, a transit orbit is observed for sufficiently large values of $ \ell $. On the other hand, for uncharged particle, all resulting orbit configurations are transit for same energy and spacetime parameters. As for case $ \bar{E}^{2} < 1 $, for a positively charged particle, while two bound and circular orbits can be formed for sufficiently small and critical values of NUT parameter (for the critical value $\ell=0.437329$ of the NUT parameter,  a circular orbit forms), only one bound orbit can be observed as value of NUT parameter increases. On the other hand, for negatively charged and uncharged particles, only one bound orbit (actually a many-world bound) is formed for all values of NUT parameter (again for same energy and spacetime parameters). Finally, for case $ \bar{E}^{2} =1 $, for motion of positively charged test particle, while bound (a many-world bound) and circular orbits can be observed for sufficiently small and critical values of NUT parameter (for the critical value $\ell=0.747708$ of the NUT parameter, again a circular orbit forms), a crossover escape orbit (for region $ - \infty < r < r_{1}$, $r_{1}$ being the root of radial polynomial) can be formed for sufficiently large values of $ \ell $. On the other hand, for a negatively charged particle, while one can observe one bound and escape orbits for small values of NUT parameter, one can encounter crossover escape orbits (for region $ r_{1} < r < \infty $) for sufficiently large values of NUT parameter. As for an uncharged particle, for same spacetime parameters chosen, only crossover escape orbits are observed for all values of NUT parameter.

\subsection{Existence of bound orbits in causality-preserving region}

It is obvious that  the particle moves in a bound orbit  if the radial motion is constrained in the interval $r_2 \leq r \leq r_1$ ($r_2$ and $r_1$ are finite) where $r_2$ and $r_1$ correspond to turning points of radial potential $\bar{P}_r(r)$ such that $\bar{P}_r(r_2)=\bar{P}_r(r_1)=0$. From physical point of view, it will be of great interest to investigate existence of bound orbits in causality-preserving region. As discussed in \cite{chandrasekhar}, extending spacetime to allow negative $r$ values, it is certain that causality is violated in the domain for which $ g_{\varphi \varphi}<0$ and $\varphi$ becomes a time-like coordinate. Hence, the domains for which $ g_{\varphi \varphi} > 0$ and space-like character of $ \varphi $-coordinate is preserved can be identified as causality-preserving regions. For KNTN spacetime, it is obvious from metric structure that $ \left. g_{\varphi \varphi}\right|_{\theta=\frac{\pi}{2}} >0$ for $r>r_+$ which implies that the region where $r>r_+ $ is a causality-preserving region. Of course, for KNTN spacetime, there exist other regions where metric component $ \left. g_{\varphi \varphi}\right|_{\theta=\frac{\pi}{2}} > 0 $. However, as done in \cite{wilkins}, we concentrate on existence of bound orbits in the causality-preserving region where $r>r_+$ (i.e outside outer singularity). In what follows, we will investigate conditions for which a bound orbit exists in region where $ r > r_+ $ for  $\bar{E}^2>1$,  $\bar{E}^2<1$ and  $\bar{E}^2=1$. We should remark that it has been shown in \cite{wilkins} for Kerr spacetime that  for $\bar{E}^2>1$ there exist no bound orbit in causality preserving region (bound orbit in region where $r>r_+$). To make an analysis of such a bound motion for our spacetime, we follow a similar procedure outlined in \cite{wilkins}. In this sense, the conditions for the existence of bound orbits can be determined by using Descartes' rule of sign. According to Descartes' rule of sign, a polynomial possessing real coefficients can not have more positive roots than the number of variations of sign in its coefficients. First of all, the conditions that we will obtain will imply that a radial bound interval $r_2 \leq  r \leq r_1$ may exist such that for that region, $\bar{P}_r(r)>0$ and $ r_2>r_+ $. Now following a similar procedure as done in \cite{wilkins}, we can affect a transformation $r=R+ r_+$, where $r_+$ describes the metric singularity  (i.e. $\Delta(r_+)=0$), assuming that at least one region of binding exists where $r>r_+$. To this end, we express  $\bar{P}_r(r)$ in terms of new variable $R$. Then in terms of $R$, we obtain
\begin{equation}
\bar{P}_R(R)=A_4 R^4 + A_3 R^3+A_2 R^2 +A_1 R +A_0
\end{equation}
where
\begin{equation}
A_0=\frac{1}{4 \bar{E}^2} \left[ a^2+2 \bar{E} \left( \ell^2 \bar{E}+r_+ (\bar{E} r_+-\bar{q} Q)\right)\right]^2,
\end{equation}
\begin{eqnarray}
A_1&=& \frac{a^2}{2 \bar{E}^2} \left[ M-2\bar{q} Q \bar{E}+r_+ (4 \bar{E}^2-1)\right]+2 \ell^2 \left[ M-\bar{q} Q \bar{E}+r_+(2 \bar{E}^2-1)\right]\nonumber\\
& & +2r_+\left[ \bar{q}^2 Q^2+r_+ (M-3 \bar{q} Q \bar{E})+r_+^2(2 \bar{E}^2-1)\right],
\end{eqnarray}
\begin{eqnarray}
A_2&=&\bar{q}^2 Q^2+a^2 \left( 1-\frac{1}{4 \bar{E}^2}\right)+\ell^2 (2 \bar{E}^2-1)\nonumber\\
&& +2 r_+(2M-3\bar{q}Q\bar{E})+r_+^2(6\bar{E}^2-5),
\end{eqnarray}
\begin{equation}
A_3=2 \left[ M-\bar{q} Q \bar{E}+2r_+(\bar{E}^2-1)\right]
\end{equation}
and
\begin{equation}
A_4=\bar{E}^2-1.
\end{equation}
At this stage, let us consider in what conditions this polynomial has positive roots (which will lead to the existence of bound orbit for $r>r_+$). We remark that $A_0>0$.  Then according to Descartes' rule of sign if $A_4>0$ ($\bar{E}^2 >1$), a bound region for $r>r_+$ may be realised under the conditions
\begin{equation}
A_1<0, \qquad A_2>0, \qquad A_3<0 \label{interval_bound_1}
\end{equation}
since we also have $A_0>0$. Then four variations of sign would be possible and therefore there may exist four (distinct) real positive roots for $\bar{P}_R(R)$. If the above inequality conditions are simultaneously met, we have the possibility of having a bound motion for $\bar{E}^2 >1$ in the region where $r>r_+$.  Such a bound orbit for $\bar{E}>1$ and $r>r_+$ can be realised for the parameters $\bar{E}=1.1$, $M=1$, $Q=5$, $\ell=6$, $a=3.4$ and $\bar{q}=5$. With these parameters, the conditions  (\ref{interval_bound_1}) are simultaneously met and therefore an interval bound region exists in the region where $r_+<r_3<r<r_2$ ($r_2$ and $r_3$ corresponds to zeros of radial potential $\bar{P}_r(r)$). As a further remark, we should state that for an uncharged particle ($\bar{q}=0$), there exist no bound region for $r>r_+$ since all $A_i$'s ($i=1,2,3,4$) will be positive for $\bar{q}=0$. Furthermore, as can be seen from the coefficients $A_i$, we should also remark that for $\bar{E}^2>1$ and $\bar{q} Q \bar{E}<0$, all $A_i$'s become positive such that there would be no sign change for the polynomial $\bar{P}_R(R)$. Therefore $\bar{P}_r(r)$ will not possess (real) roots for $r>r_+$ and as a result there would be no bound orbit for $r>r_+$ for the case where $\bar{E}^2>1$. It means that, for a test particle with positive energy ($\bar{E}>1$) but possessing electric charge $\bar{q}$ with opposite sign (i.e $\bar{q}<0$, $Q>0$), a bound orbit in causality-preserving region cannot form in a black hole spacetime with NUT charge. Similarly, a bound orbit in such a region cannot exist in a black hole spacetime if the test particle with charge  $\bar{q}$ having same sign with charge $Q$ of spacetime possesses negative energy ($\bar{E}<-1$). With a similar reasoning, if the condition $M>3 \bar{q}Q \bar{E}$ (with $\bar{q} Q \bar{E}>0$) holds, again a bound orbit cannot form in the causality-preserving region where $r>r_+$.   

\noindent On the other hand, if $A_4<0$ ($\bar{E}^2 <1$), there exist at most three variations of sign since $A_0>0$. For this case, then either of the following  inequalities should be simultaneously fulfilled for the existence of bound region(s):
\begin{eqnarray}
A_1<0, \qquad A_2<0, \qquad A_3>0, \nonumber\\
A_1>0, \qquad A_2<0, \qquad A_3>0, \nonumber\\
A_1<0, \qquad A_2>0, \qquad A_3>0, \nonumber\\
A_1<0, \qquad A_2>0, \qquad A_3<0. \nonumber
\end{eqnarray}
Similarly, this implies that if any one of the above conditions are simultaneously met, there may exist at most one region of binding outside the outer singularity where $r>r_+$. 

\noindent On the other hand, for $\bar{E}^2=1$ ($A_4=0$), the existence of such orbits for $r>r_+$ depends on the sign of $A_3$. For a third order polynomial, there should be at most three variations of sign in order to obtain a bound orbit for $r>r_+$. Then, one can conclude that, if $A_3>0$ ($M>\bar{q}Q$) there would be at most two variations of sign (since $ A_{0} > 0 $) and therefore no bound orbit is seen for $r>r_+$. On the other hand, if $A_3<0$ ($M<\bar{q}Q$) and the conditions $A_2>0$, $A_1<0$ are simultaneously met, there would be three variations of sign and therefore a bound motion can be realised in the region where $r>r_+$. 

\subsection{Existence of circular orbits}

\noindent In this section, we investigate the existence of circular orbits and determine required conditions for the existence of them at  $r_c=0$ and $r_c \neq 0$. It is clear that, when the conditions
\begin{equation}
\bar{P}_{r} (r_c)=0, \qquad \left.\frac{d \bar{P}_r}{d r}\right|_{r=r_c}=0\label{circular_1}
\end{equation}
are satisfied, one gets a circular orbital motion over the equatorial plane. These two conditions require that
\begin{eqnarray}
&&(\bar{E}^2-1) r_c^4+2(M-\bar{q}Q\bar{E})r_c^3+\left(2 \bar{E}^2 \ell^2+Q^2 (\bar{q}^2-1)-\frac{a^2}{4 \bar{E}^2}\right) r_c^2\nonumber\\
&&+\left(2 (M-\bar{q} Q \bar{E})  \left(  \ell^2+\frac{a^2}{4 \bar{E}^2} \right)-  \frac{\bar{q} Q a^2}{2 \bar{E}}\right)r_c\nonumber\\
&&+\bar{E}^2 \ell^4+(\ell^2-Q^2) \left( \ell^2+\frac{a^2}{4 \bar{E}^2}\right)=0, \label{circular_2}
\end{eqnarray}
\begin{eqnarray}
&&4(\bar{E}^2-1) r_c^3+6(M-\bar{q}Q\bar{E})r_c^2+2\left(2\bar{E}^2 \ell^2+ Q^2(\bar{q}^2-1)-\frac{a^2}{4 \bar{E}^2}\right)r_c\nonumber\\
&&+2 (M-\bar{q} Q \bar{E})  \left(  \ell^2+\frac{a^2}{4 \bar{E}^2} \right)-  \frac{\bar{q} Q a^2}{2 \bar{E}}=0. \label{circular_3}
\end{eqnarray}
We notice that the former (i.e. equation (\ref{circular_2})) can also be written in the equivalent form

\begin{eqnarray}
&& 3 (\bar{E}^2-1)r_c^4+4 (M-\bar{q} Q \bar{E}) r_c^3+ \left(2 \bar{E}^2 \ell^2 + Q^2 (\bar{q}^2-1)-\frac{a^2}{4 \bar{E}^2} \right) r_c^2 \nonumber\\
&& -\bar{E}^2 \ell^4-(\ell^2-Q^2) \left( \ell^2+\frac{a^2}{4 \bar{E}^2} \right)=0. \label{circular_4}
\end{eqnarray}\\
In what follows, we try to obtain physically acceptable conditions for the existence of circular orbits at $ r_{c} = 0 $ and $ r_{c} \neq 0 $. \\

\noindent {\bf i}. Existence of circular orbits at $r_c=0$ :\\

\noindent It is obvious from the equations (\ref{circular_2}) and (\ref{circular_3}), if the conditions
\begin{eqnarray}
\bar{E}^2 \ell^4+(\ell^2-Q^2) \left( \ell^2+\frac{a^2}{4 \bar{E}^2} \right)=0, \label{circular_4a}\\ 2 (M-\bar{q} Q \bar{E})  \left(  \ell^2+\frac{a^2}{4 \bar{E}^2} \right)-  \frac{\bar{q} Q a^2}{2 \bar{E}}=0 \nonumber
\end{eqnarray}
are simultaneously satisfied, a circular orbit exists at the point $r_c=0$. Then it is clear from (\ref{circular_4a}) that irrespective of spacetime parameters, a circular orbit at $ r_c = 0 $ cannot form for an uncharged test particle. On the other hand, expressions above suggest that for a charged test particle if condition $\ell \geq Q$ holds between spacetime parameters, a circular orbit at $ r_{c} = 0 $ again cannot exist.\\  

\noindent {\bf ii}. Existence of circular orbits for $r_c \neq 0$ :\\

\noindent Now, we discuss the conditions for existence of circular orbits for $ r_{c} \neq 0 $. Before making such an analysis, we find it useful to remark that, the radial potential $\bar{P}_r(r)$ can also be expressed in the form 
\begin{equation}
\bar{P}_r(r)=(r-r_c)^2 \left( (\bar{E}^2-1) r^2+\mu_1 r+\mu_2\right)\label{analysis_10}
\end{equation}
where
\begin{equation}
\mu_1= 2 \left((\bar{E}^2-1) r_c + M-\bar{q} Q \bar{E}\right), \label{analysis_11}
\end{equation}
\begin{equation}
\mu_2= 3 (\bar{E}^2-1) r_c^2 + 4 \left( M-\bar{q} Q \bar{E}\right)r_c+2 \bar{E}^2 \ell^2+Q^2(\bar{q}^2-1)-\frac{a^2}{4 \bar{E}^2} \label{analysis_12}
\end{equation}
 (i.e. $\bar{P}_r(r)$ has one double zero at $r=r_c$) for which the conditions (\ref{circular_1}) are respected. It is obvious that when the extra condition 
\begin{equation}
\mu_1^2-4(\bar{E}^2-1)\mu_2=0
\end{equation}
is satisfied, $\bar{P}_r(r)$ can be written in the form
\begin{equation}
\bar{P}_r(r)=(\bar{E}^2-1)(r-r_c)^2 (r-r_c^{\prime} )^2
\end{equation}
which implies that  there exist another circular orbit at $r=r_c^{\prime} $ (in addition to circular orbit at $r=r_c$) where $r_c^{\prime} $ can be calculated as
\begin{equation}
r_c^{\prime} =\frac{(\bar{q} Q \bar{E}-M)}{\bar{E}^2-1}-r_c \,  , \qquad \bar{E}^2\neq 1.
\end{equation}
\noindent Now, our aim is to make an analysis of equations (\ref{circular_3}) and (\ref{circular_4}) to investigate conditions for existence of circular orbits at $r_c \neq 0 $. 

\noindent First, we should note that, in most of the works in literature (as in \cite{stuchlik_1,stuchlik_2}), the conditions for the existence of circular orbit are analytically obtained for energy $\bar{E}$ and angular momentum $\bar{L}$ of the test particle. However, in our case, the angular momentum $\bar{L}$ of the test particle has already been constrained to be given by expression (\ref{motion_22}) for existence of equatorial motion in a spacetime with NUT charge. Therefore, looking at equations (\ref{circular_3}) and (\ref{circular_4}), these equations involve  $\bar{E}_{c}$, $r_c$, $\bar{q}$ and spacetime parameters. Considering a test particle with charge $\bar{q}$ fixed moving in a spacetime with parameters $M$, $Q$, $\ell$ and $a$ also kept fixed, one can see that equations (\ref{circular_3}) and (\ref{circular_4}) should be solved for $\bar{E}_{c}$ and $r_c$ for a physical analysis of circular orbits. However, we note that these equations are fourth order in both $\bar{E}_c$ and $r_c$ and therefore a simultaneous exact analytical solution of equations (\ref{circular_3}) and (\ref{circular_4}) for energy $\bar{E}_{c}$ and circular radius $r_c$ in terms of the charge $\bar{q}$ and spacetime parameters do not seem to be possible. 

\noindent On the other hand, looking at the equations, one can see that equation (\ref{circular_3}) is a linear equation in $a^2$ and $\ell^2$ while (\ref{circular_4}) is a second order equation in $\ell^2$. Then, if one eliminates $a^2$ from (\ref{circular_3}) and substitute the resulting equation into (\ref{circular_4}), one obtains a second order equation in $\ell^2$. It means that, equations (\ref{circular_3}) and (\ref{circular_4}) can simultaneously be solved analytically for $\ell^2$ and $a^2$ yielding 

\begin{eqnarray}
a^2&=&\frac{4 \bar{E}_c^2}{(r_c-M+2 \bar{E}_c \bar{q} Q)} \left[ \left(2 \bar{E}_c^2 \ r_c+M-\bar{E}_c \bar{q} Q\right) \ell^2 \right.\nonumber\\
&+&\left. 2 (\bar{E}_c^2-1)r_c^3+3(M-\bar{q} Q \bar{E}_c) r_c^2+Q^2(\bar{q}^2-1)r_c\right] \label{circular_5}
\end{eqnarray}
where
\begin{eqnarray}
\ell^2 &=&\frac{1}{2\left[ \left(2\bar{E}_c^2+1 \right)\left(\bar{q} Q \bar{E}_c +r_c\right)+\bar{E}_c^2 (r_c -M)\right]} \left[\bar{q} Q(\bar{E}_c Q^2 - \bar{q} Q r_c )\right.\label{circular_6}\\
&+&2r_c \left(\bar{E}_c^2+1 \right)\left(Q^2+r_c(r_c -M + 2 \bar{q}Q \bar{E}_c ) \right)-2r_c^2 \left(2 \bar{E}_c^2 r_c +M \right)\nonumber
\\
&\mp& \left.  \left(2 \bar{E}_c r_c (r_c -M + 2 \bar{q}Q \bar{E}_c )+\bar{q} Q r_c+ \bar{E}_c Q^2
 \right) \sqrt{(\bar{q} Q -2r_c\bar{E}_c)^2+ 4r_c(M-r_c)}\right]. \nonumber
\end{eqnarray}

\noindent Provided that the right hand sides of the expressions (\ref{circular_5}) and (\ref{circular_6}) are positive, these relations can be regarded as the required conditions for the existence of circular orbits for a charged test particle. In addition, there also exists a reality condition given by the inequality relation 
\begin{equation}
(\bar{q} Q -2r_c\bar{E}_c)^2+ 4r_c(M-r_c) \geq 0  \label{circular_6a}
\end{equation}
which makes the expression inside the square root in (\ref{circular_6}) positive. We should remark that for circular orbit conditions (\ref{circular_5}) and (\ref{circular_6}),  NUT and rotation parameters cannot be considered as a function of $r_c$ and $\bar{E}_c$ since it is obvious that test particle moves in a spacetime with parameters $a$, $M$, $\ell$, $Q$ fixed. These two conditions imply that  a test particle with charge $\bar{q}$ can move in a circular orbit in a spacetime with fixed parameters such that the energy $\bar{E}_c$ of the particle and the circular radius $r_c$  are related to each other with the conditions given by (\ref{circular_5}) and (\ref{circular_6}).

\noindent Next, let us discuss the stability of circular orbits whose existence is determined by the expressions (\ref{circular_5}) and (\ref{circular_6}). It is clear that if the inequality
\begin{equation}
\left.\frac{d^2P_r}{dr^2}\right|_{r=r_c}<0 \label{circular_7}
\end{equation}
holds, one can obtain stable circular orbits for which the stability condition reads
\begin{equation}
12(\bar{E}_c^2-1) r_c^2+12(M-\bar{q}Q\bar{E}_c)r_c+\left(4\bar{E}_c^2 \ell^2+ 2Q^2(\bar{q}^2-1)-\frac{a^2}{2 \bar{E}_c^2}\right)<0 . \label{circular_8}
\end{equation}

\noindent It would also be of great interest to examine the circular orbits for an uncharged test particle ($\bar{q}=0$). Then by solving equations (\ref{circular_3}) and (\ref{circular_4}) for an uncharged particle, one can analytically obtain the energy of the test particle as
\begin{equation}
\bar{E}_c^2=\frac{\alpha(r_c)+ \sqrt{ \alpha^2(r_c) + a^2 (r_c^2 + \ell^2)(3 r_c^2- \ell^2)(r_c^2+ \ell^2 - Q^2)}}{2(r_c^2+\ell^2)(3 r_c^2-\ell^2)} \label{circular_9}
\end{equation}
where $r_c$ should obey the relation
\begin{eqnarray}
&&2 r_c^2 \alpha(r_c)-3 (r_c^2-\ell^2) \beta(r_c)-(3 r_c^2-\ell^2)\sqrt{\beta^2(r_c)+2 a^2 r_c^3(r_c^2+\ell^2)(r_c-M)}\nonumber\\
&&+2 r_c^2 \sqrt{\alpha^2(r_c)+a^2(r_c^2+\ell^2)(3 r_c^2-\ell^2)(r_c^2+\ell^2-Q^2)}=0 \label{circular_10}
\end{eqnarray}
as well. Here, we define
\begin{equation}
\alpha(r_c)=(r_c^2+\ell^2)^2+(r_c^2-\ell^2) (2 r_c^2+\ell^2)-4 M r_c^3 \label{circular_11}
\end{equation}
and
\begin{equation}
\beta(r_c)=r_c^2(2 r_c^2+Q^2)-(3 r_c^2+\ell^2) M r_c . \label{circular_12}
\end{equation}

\noindent Obviously, the existence of circular orbits for an uncharged particle ($\bar{q}=0$) depends on the positivity of the right hand side of the equation (\ref{circular_9}). It should also be noted that, equation (\ref{circular_10}) satisfied by circular radius can be solved (at least numerically) in terms of spacetime parameters. If a physical solution exists for $ r_{c} $, then from (\ref{circular_9}), one can obtain the energy $ \bar{E}_{c} $ of an uncharged test particle in terms of spacetime parameters as well. Nevertheless, an exact analytical solution does not seem to be possible. 

\subsection{Equatorial Newtonian orbits}

\noindent To further exploit the physical effect of the NUT parameter on the motion over equatorial plane, we make an analysis of the Newtonian orbits as well, where the radial variable $r$ for those orbits is assumed to be much larger than the Schwarzschild radius of the gravitational source ($r >> r_S=2M$).  For that, we analyse the orbit equation (\ref{motion_25}) over the equatorial plane and utilise the physically oriented approximations raised in \cite{dereli}. First, one can express the orbit equation (\ref{motion_25}) in terms of a new variable $u$ such that $r=\frac{1}{u}$. With this substitution, the orbital equation (\ref{motion_25}) turns into
\begin{equation}
\left( \frac{du}{d \varphi} \right)^2 = \frac{\Delta_u^2 \left(  \left[ 2 (\ell^2 u^2+1)\bar{E}^2 +a^2 u^2  -2 \bar{q} Q \bar{E} u  \right]^2 - \Delta_u \left( 4 \bar{E}^2 ( \ell ^2 u^2 +1)+ a^2 u^2 \right) \right)}
{a^2 \left( 2(\ell^2 u^2+1) \bar{E}^2 + a^2 u^2 - \Delta_u -2 \bar{q} Q \bar{E} u \right)^2} \label{newtonian_1}
\end{equation}
where we define
\begin{equation}
\Delta_u= 1- 2 M u + (a^2 -\ell^2 +Q^2 )u^2. \label{newtonian_2}
\end{equation}

\noindent Thus, if one assumes that the radius of a Newtonian orbit is much larger than the corresponding Schwarzschild radius of the gravitational source, the orbital equation (\ref{newtonian_1}) may be expanded around $u = 0$ up to third order in order to compare its relativistic corrections to Newtonian orbits with those in a Schwarzschild background \cite{chandrasekhar}: 
\begin{equation}
\left( \frac{du}{d \varphi}\right)^2\simeq f(u) = D_0 +D_1 u +D_2 u^2 + D_3 u^3, \label{newtonian_3}
\end{equation}
where
\begin{equation}
 D_0=\frac{4 \bar{E}^2 (\bar{E}^2-1)}{a^2(2 \bar{E}^2-1)^2} , \label{newtonian_4}
\end{equation}

\begin{eqnarray}
 D_1= \frac{16 (2M\bar{E}- \bar{q} Q)(1-\bar{E}^2)\bar{E}^3}{a^2(2 \bar{E}^2-1)^3}
 + \frac{8(M-\bar{q} Q \bar{E})\bar{E}^2}{a^2(2 \bar{E}^2-1)^2} , \label{newtonian_5}
\end{eqnarray}

\begin{eqnarray}
 D_2&=& \frac{48 \left(\bar{E}^2-1\right) (\bar{q} Q-2 M \bar{E})^2 \bar{E}^4}{a^2(2 \bar{E}^2-1)^4}   \nonumber \\
   && + \frac{16 \bar{E} ^4}{a^2(2 \bar{E}^2-1)^3}\left[2M \bar{E} (3\bar{q}Q -2M \bar{E} )-2\bar{q}^2 Q^2  + (\bar{E}^2-1)(Q^2 -2 \ell^2)\right] \nonumber \\
   & & + \frac{4 \bar{E}^2}{a^2(2 \bar{E}^2-1)^2} \left[3 a^2 (\bar{E}^2-1) + Q^2 \left(\bar{q}^2-1\right) + 2 \ell^2 \bar{E}^2 \right]-1 ,\label{newtonian_6}
\end{eqnarray}

\begin{eqnarray}
D_3&=&\frac{128 \bar{E}^5}{a^2(2 \bar{E}^2-1)^5}(\bar{q} Q-2 M \bar{E})^3(\bar{E}^2-1)\nonumber \\
&& +\frac{96\bar{E}^4(\bar{q}Q -2M \bar{E})}{a^2(2 \bar{E}^2-1)^4} \left[(\bar{E}^2-1)(Q^2-2 \ell^2) \bar{E}-(\bar{q}Q -2M \bar{E})(\bar{E} \bar{q} Q + M(1- 2 \bar{E}^2) ) \right]\nonumber \\
&& + \frac{16 \bar{E}^3 }{a^2(2 \bar{E}^2-1)^3}\left[3 a^2 (\bar{q} Q-2M \bar{E}) (\bar{E}^2-1)+2 \bar{E}^2(\bar{q} Q-M \bar{E})(2 \ell^2-Q^2)\right.\nonumber \\
&&\left.+(\bar{q} Q-2M \bar{E}) \left( 4 M \bar{E} (M \bar{E}-\bar{q} Q)+\ell^2 (3 \bar{E}^2-1)+ Q^2(\bar{q}^2-\bar{E}^2)\right)\right]\nonumber \\
&& + \frac{8 \bar{E}^2 }{a^2(2 \bar{E}^2-1)^2}\left[(M -\bar{q} Q \bar{E}) (a^2(2 \bar{E}^2+1)+\ell^2)+M a^2(\bar{E}^2-1)\right]+2 M \label{newtonian_7}
\end{eqnarray}
provided that $2\bar{E}^2-1 \neq0$ and $a\neq0$. Now we concentrate  on elliptical type  Newtonian orbits. For these type of orbits, it requires that both $D_0$ and $D_2$ be negative provided that $D_3>0$. This can happen for $\bar{E}^2<1$. First, we should point out that classical (elliptical type) Newtonian orbits are described by the equation
\begin{equation}
\left( \frac{du}{d \varphi}\right)^2 = D_0 +D_1 u - u^2  \label{newtonian_8}
\end{equation}
(where we omit correction terms) whose solution leads to Kepler solution
\begin{equation}
u=\frac{1}{r}=\frac{1}{\hat{r}_0} (1+\epsilon \cos \phi) \label{newtonian_9}
\end{equation}
where we identify $\hat{r}_0=\frac{2}{D_1}$ as Newtonian Kepler orbit parameter and eccentricity $\epsilon=\sqrt{\frac{4 D_0}{D_1^2}+1}$.

\noindent In this sense, looking at the orbit equation (\ref{newtonian_3}), one can see that all the terms in $D_2$ except $-1$ and all the terms in $D_3$ describe general relativistic corrections (or improvements) to the classical Newtonian orbits. It is remarkable that, unlike the other spacetime parameters ($a$, $Q$ and $M$), the effect of the NUT parameter can be explicitly seen as the general relativistic correction (improvement) to the classical Newtonian orbits, where the NUT parameter contributes at least at the order $u^2$ (and the higher orders). 

\noindent At this point, we remark that rotation parameter $a$ appears at the zeroth order of orbital equation (\ref{newtonian_3}). Usually at zeroth order of orbit equation, the angular momentum $\bar{L}$ (as an independent parameter) appears as in \cite{chandrasekhar} and \cite{cebeci2}. However, the existence of equatorial orbits in a spacetime with NUT parameter requires that the angular momentum $\bar{L}$ is related to rotation parameter $a$ and energy $\bar{E}$ through the relation $\bar{L}=\frac{a(2 \bar{E}^2-1)}{2\bar{E}}$. Since we have already used this relation in Section 3 to express our equations of motion over the equatorial plane, rotation parameter $a$ appears at the zeroth order of our orbital equation. 

\noindent Before closing this section, we would like to point out that the solution of orbital equation (\ref{newtonian_3}) can be expressed in terms of Jacobian elliptic functions. Then for elliptical type orbits, following the physical arguments raised in \cite{dereli} and \cite{cebeci2} and recalling periodicity of Jacobian elliptic functions, one can obtain perihelion shift per revolution as
\begin{equation}
\hat{\Sigma} = \frac{2 \pi}{ \sqrt{|D_{2}|} } \left( 1 + \frac{3}{ 2 ( 1 - \epsilon^{2} ) \hat{r}_{0}} \frac{ D_{3} }{ |D_{2}| } \right) - 2 \pi \label{newtonian_10}  
\end{equation}
in terms of orbit parameters $ \hat{r}_{0} $ and $ \epsilon $. Here, the effect of NUT parameter can be explicitly seen through the coefficients $ D_{2} $ and $ D_{3} $.

\section{Analytical solutions of the orbit equations}
In this section, to obtain the trajectory of charged test particle in equatorial KNTN spacetime, we solve orbit equations (\ref{motion_24})-(\ref{motion_26}) analytically where $\bar{P}_r(r)$ is given by (\ref{motion_27}). The solutions of orbit equations describe the trajectory of non-geodesic motion of massive charged test particle in KNTN spacetime. These exact solutions also illustrate the effect of the NUT parameter on the equatorial trajectory of the test particle.

\subsection{The solution of  (\ref{motion_24})  :}
To obtain the exact analytical solution of the radial equation (\ref{motion_24}), one can perform the transformation (for $\bar{E}^2\neq1$)
\begin{equation}
r=\frac{\alpha_3}{\left(4 y-\frac{\alpha_2}{3}\right)}+r_1 \label{analytic_7}
\end{equation}
where $r_1$ is assumed to be one real root of $\bar{P}_r(r)$. Defining
\begin{equation}
\alpha_1=B_3+4 B_4 r_1, \label{analytic_8}
\end{equation}
\begin{equation}
\alpha_2=B_2+3 B_3 r_1+6 B_4 r_1^2, \label{analytic_9}
\end{equation}
\begin{equation}
\alpha_3=B_1+2 B_2 r_1+3 B_3 r_1^2+4 B_4 r_1^3 \label{analytic_10},
\end{equation}
where the coefficients $ B_{i} , i=0,1,2,3,4$ are given in (\ref{radial_potential_2})-(\ref{radial_potential_6}), equation (\ref{motion_24}) can be brought into the standard Weierstrass form
\begin{equation}
\left( \frac{d y}{d \lambda}\right)^2=\bar{P}_3 (y)=4 y^3-g_2 y-g_3 \label{analytic_11}
\end{equation}
whose solution can be written in terms of Weierstrass $\wp$ function \cite{wang}
\begin{equation}
y(\lambda)=\wp((\lambda-\lambda_0); g_2, g_3) \label{analytic_12}
\end{equation}
with 
\begin{equation}
g_2=\frac{1}{12} \left( \alpha_2^2-3 \alpha_1 \alpha_3\right), \qquad g_3=\frac{1}{8} \left( \frac{\alpha_1 \alpha_2 \alpha_3}{6}-\frac{B_4 \alpha_3^2}{2}-\frac{\alpha_2^3}{27}\right). \label{analytic_13}
\end{equation}
Then, the solution for radial coordinate $r$ can be given by
\begin{equation}
r=\frac{\alpha_3}{\left[ 4 \wp \left( (\lambda-\lambda_0); g_2, g_3\right)-\frac{\alpha_2}{3}\right]} +r_1 . \label{analytic_14}
\end{equation}

\subsection{The solution of  (\ref{motion_25})  :}
\noindent Next, from the integration of (\ref{motion_25}), one obtains
\begin{equation}
\frac{1}{a} (\varphi- \varphi_0)= \int^r \frac{\left( \bar{E}(r^2+\ell^2) +\frac{a^2}{2 \bar{E}}- \bar{q}Qr - \frac{\Delta(r)}{2 \bar{E}}\right)}{\Delta(r) \sqrt{\bar{P}_r(r)}} dr . \label{analytic_15}
\end{equation}

\noindent Using the remark that $ \int^r \frac{dr}{\sqrt{\bar{P}_r(r)}}=\int^y \frac{dy}{\sqrt{\bar{P}_3(y)}}= \lambda- \lambda_0$, one can accomplish the integration of the right hand side with respect to radial coordinate $r$ resulting in
\begin{eqnarray}
\frac{1}{a} (\varphi- \varphi_0) &=&  \left[  \left( \bar{E} \ell^2 + \frac{1}{2 \bar{E}} \left( \ell^2- Q^2\right) \right) \omega_0 
+  \left( \frac{M}{\bar{E}}- \bar{q} Q \right) \bar{\omega}_0 + \left( \frac{2 \bar{E}^2 -1}{2\bar{E}}\right)  \tilde{\omega}_0
 \right](\lambda- \lambda_0) \nonumber \\
& & + \sum_{i=1}^2  \sum_{j=1}^2   \left[  \left( \bar{E} \ell^2 + \frac{1}{2 \bar{E}} \left( \ell^2- Q^2\right) \right) \omega_i 
+  \left( \frac{M}{\bar{E}}- \bar{q} Q \right) \bar{\omega}_i+ \left( \frac{2 \bar{E}^2 -1}{2\bar{E}}\right)  \tilde{\omega}_i
 \right] \nonumber \\
 & &  \times \frac{1}{ \wp^{\prime}  (y_{ij})}  \left[ \zeta (y_{ij}) (\lambda-\lambda_0) + \ln \left(\frac{\sigma(s-y_{ij})}{\sigma(s_0-y_{ij})} \right)\right] . \label{analytic_16}
\end{eqnarray}
Here, $\wp(y_{ij})=y_i$ with $ \wp(y_{11})=\wp(y_{12})=y_1$,  $ \wp(y_{21})=\wp(y_{22})=y_2$ where
\begin{equation}
y_1=\frac{1}{4 \Delta(r_1)}\left( \frac{\alpha_2}{3} \Delta(r_1)-\alpha_3 r_1+\alpha_3 r_-\right) , \label{analytic_17}
\end{equation}

\begin{equation}
y_2=\frac{1}{4 \Delta(r_1)}\left( \frac{\alpha_2}{3} \Delta(r_1)-\alpha_3 r_1+\alpha_3 r_+\right), \label{analytic_18}
\end{equation}
and the variables $s$ and $\lambda$ are related by $ s-s_0= \lambda-\lambda_0$, $ s_0$ and $\lambda_0$ being integration constants. We further identify
\begin{equation}
\omega_0= \frac{1}{\Delta(r_1)}, \label{analytic_19}
\end{equation}

\begin{equation}
\omega_1= -\frac{\alpha_3(r_1-r_-)^2}{4 \Delta^2(r_1)(r_+-r_-)}, \label{analytic_20}
\end{equation}

\begin{equation}
\omega_2= \frac{\alpha_3(r_1-r_+)^2}{4 \Delta^2(r_1)(r_+-r_-)}, \label{analytic_21}
\end{equation}

\begin{equation}
\bar{\omega}_0=  \frac{r_1}{\Delta(r_1)}, \label{analytic_22}
\end{equation}

\begin{equation}
\bar{\omega}_1= \frac{\alpha_3 (r_1-r_-)}{4 \Delta^2(r_1)(r_+-r_-)} \left[ \Delta(r_1)+r_1(r_--r_1)\right], \label{analytic_23}
\end{equation}

\begin{equation}
\bar{\omega}_2= \frac{\alpha_3 (r_+-r_1)}{4 \Delta^2(r_1)(r_+-r_-)} \left[ \Delta(r_1)+r_1(r_+-r_1)\right], \label{analytic_24}
\end{equation} 

\begin{equation}
\tilde{\omega}_0=  \frac{r_1^2}{\Delta(r_1)}, \label{analytic_25}
\end{equation} 

\begin{equation}
\tilde{\omega}_1= -\frac{\alpha_3}{4 \Delta^2(r_1)(r_+-r_-)} \left[ \Delta(r_1)+r_1(r_--r_1)\right]^2, \label{analytic_26}
\end{equation}

\begin{equation}
\tilde{\omega}_2= \frac{\alpha_3}{4 \Delta^2(r_1)(r_+-r_-)} \left[ \Delta(r_1)+r_1(r_+-r_1)\right]^2. \label{analytic_27}
\end{equation}

\subsection{The solution of  (\ref{motion_26})  :}
\noindent Finally, the integration of (\ref{motion_26}) yields
\begin{equation}
t-t_0 = \int^r \frac{\left[(r^2+a^2+\ell^2)\left(\bar{E}(r^2+\ell^2)+\frac{a^2}{2 \bar{E}}-\bar{q} Q r\right)-\frac{a^2 \Delta(r)}{2 \bar{E}}\right]}{\Delta(r) \sqrt{\bar{P}_r(r)}} dr \label{analytic_28}
\end{equation}
where upon integration, one can obtain the result 
\begin{eqnarray}
t-t_0=&& \left[ \left( \bar{E} \ell^2 (a^2+\ell^2)+\frac{a^2 \ell^2}{\bar{E}}-\frac{a^2 Q^2}{2 \bar{E}}\right)\omega_0+\left( \frac{Ma^2}{\bar{E}}-(a^2+\ell^2)\bar{q}Q\right)\bar{\omega}_0\right.\nonumber\\
&&\left.+\bar{E}(a^2+2 \ell^2)\tilde{\omega}_0-\bar{q}Q\hat{\omega}_0+\bar{E}\check{\omega}_0\right](\lambda-\lambda_0)\nonumber\\
&&+\sum_{i=1}^3  \sum_{j=1}^2 \frac{(\bar{E}\check{\omega}_i-\bar{q}Q\hat{\omega}_i)}{ \wp^{\prime}  (y_{ij})}  \left[ \zeta (y_{ij}) (\lambda-\lambda_0) + \ln \left(\frac{\sigma(s-y_{ij})}{\sigma(s_0-y_{ij})}\right) \right] \nonumber\\
&&+\sum_{i=1}^2  \sum_{j=1}^2 \left[ \left( \bar{E}\ell^2(a^2+\ell^2)+\frac{a^2\ell^2}{\bar{E}}-\frac{a^2Q^2}{2\bar{E}}\right)\omega_i\right.\nonumber\\
&&\left.+\left( \frac{a^2 M}{\bar{E}}-(a^2+\ell^2)\bar{q}Q\right)\bar{\omega}_i+\bar{E}(a^2+2\ell^2)\tilde{\omega}_i\right]\nonumber\\
&&\times\frac{1}{ \wp^{\prime}  (y_{ij})}  \left[ \zeta (y_{ij}) (\lambda-\lambda_0) + \ln \left(\frac{\sigma(s-y_{ij})}{\sigma(s_0-y_{ij})}\right) \right] \nonumber\\
&&-\bar{E} \check{\omega}_4 \sum_{j=1}^2 \frac{1}{\wp^{\prime \, 2}(y_{3j})}\left[ \left( \wp(y_{3j})+\frac{\wp^{\prime \prime}(y_{3j})}{\wp^{\prime}(y_{3j})}\right)(\lambda-\lambda_0)\right.\nonumber\\
&&+\left.\left( \zeta(s-y_{3j})+\frac{\wp^{\prime \prime}(y_{3j})}{\wp^{\prime}(y_{3j})}\ln \left(\frac{\sigma(s-y_{3j})}{\sigma(s_0-y_{3j})}\right)-\zeta_0^{(j)}\right) \right]. \label{analytic_29}
\end{eqnarray}
In addition, we identify $\wp(y_{3j})=y_3=\frac{\alpha_2}{12}  \ (j=1,2)$ with  $ \wp(y_{31})=\wp(y_{32})=y_3=\frac{\alpha_2}{12}$. We further calculate 

\begin{equation}
\hat{\omega}_0= \frac{r_1^3}{\Delta(r_1)}, \label{analytic_30}
\end{equation}

\begin{equation}
\hat{\omega}_1= \frac{\alpha_3\left[ \Delta(r_1)+r_1(r_--r_1)\right]^3}{4 \Delta^2(r_1)(r_+-r_-)(r_1-r_-)}, \label{analytic_31}
\end{equation}

\begin{equation}
\hat{\omega}_2= \frac{\alpha_3\left[ \Delta(r_1)+r_1(r_+-r_1)\right]^3}{4 \Delta^2(r_1)(r_+-r_-)(r_+-r_1)}, \label{analytic_32}
\end{equation}

\begin{equation}
\hat{\omega}_3= \frac{\alpha_3 \Delta(r_1)}{4 (r_1-r_-)(r_1-r_+)}, \label{analytic_33}
\end{equation}

\begin{equation}
\check{\omega}_0=\frac{r_1^4}{\Delta(r_1)}, \label{analytic_34}
\end{equation}

\begin{equation}
\check{\omega}_1=\frac{\Delta^2(r_1) r_1^4 \alpha_2^2 \left[ \Delta(r_1) \alpha_2+3(r_--r_1) \alpha_3\right]^2+81\left[ \Delta(r_1)+(r_--r_1) r_1\right]^4 \alpha_3^4}{324 \Delta^2(r_1)(r_--r_+)(r_--r_1)^2 \alpha_3^3}, \label{analytic_35}
\end{equation}

\begin{equation}
\check{\omega}_2=\frac{\Delta^2(r_1) r_1^4 \alpha_2^2 \left[ \Delta(r_1) \alpha_2+3(r_+-r_1) \alpha_3\right]^2+81\left[ \Delta(r_1)+(r_+-r_1) r_1\right]^4 \alpha_3^4}{324 \Delta^2(r_1)(r_+-r_-)(r_+-r_1)^2 \alpha_3^3}, \label{analytic_36}
\end{equation} 

\begin{equation}
\check{\omega}_3=\frac{\Delta(r_1)\left[ r_1^4 \alpha_2^3 \left( \Delta(r_1)(M-r_1) \alpha_2+3(r_--r_1)(r_+-r_1)\alpha_3\right)+81 \Delta(r_1)(r_1+M) \alpha_3^4\right]}{162(r_1-r_-)^2(r_1-r_+)^2 \alpha_3^3}, \label{analytic_37}
\end{equation}

\begin{equation}
\check{\omega}_4=\frac{ \Delta(r_1)\left( r_1^4 \alpha_2^4 + 81 \alpha_3^4 \right)}{1296 (r_1-r_-)(r_1-r_+) \alpha_3^2}. \label{analytic_38}
\end{equation}

\noindent Before closing this section, again we should point out that these solutions have been obtained for non-zero NUT parameter where the relation (\ref{motion_22}) has been used as a constraint between energy and angular momentum of test particle moving over the equatorial plane of KNTN spacetime. To examine limiting cases, we note that when $\ell=0$ (vanishing NUT parameter), the analytical solutions presented in \cite{hackmann1}  are recovered provided that if one takes $ \bar{L}=\frac{a(2 \bar{E}^2-1)}{2 \bar{E}}$ for the angular momentum and $\frac{K}{m^2}=\frac{a^2}{4 \bar{E}^2}$ for Carter constant in the related paper (by taking $\theta= \frac{\pi}{2}$ and magnetic charge to be zero as well in that work).

\begin{figure*}[htp]
    \centering
\subfloat[]
     {\label{example_of_bound_orbit_energy_greater_than_one_3_1}
\includegraphics[width=0.6\linewidth]{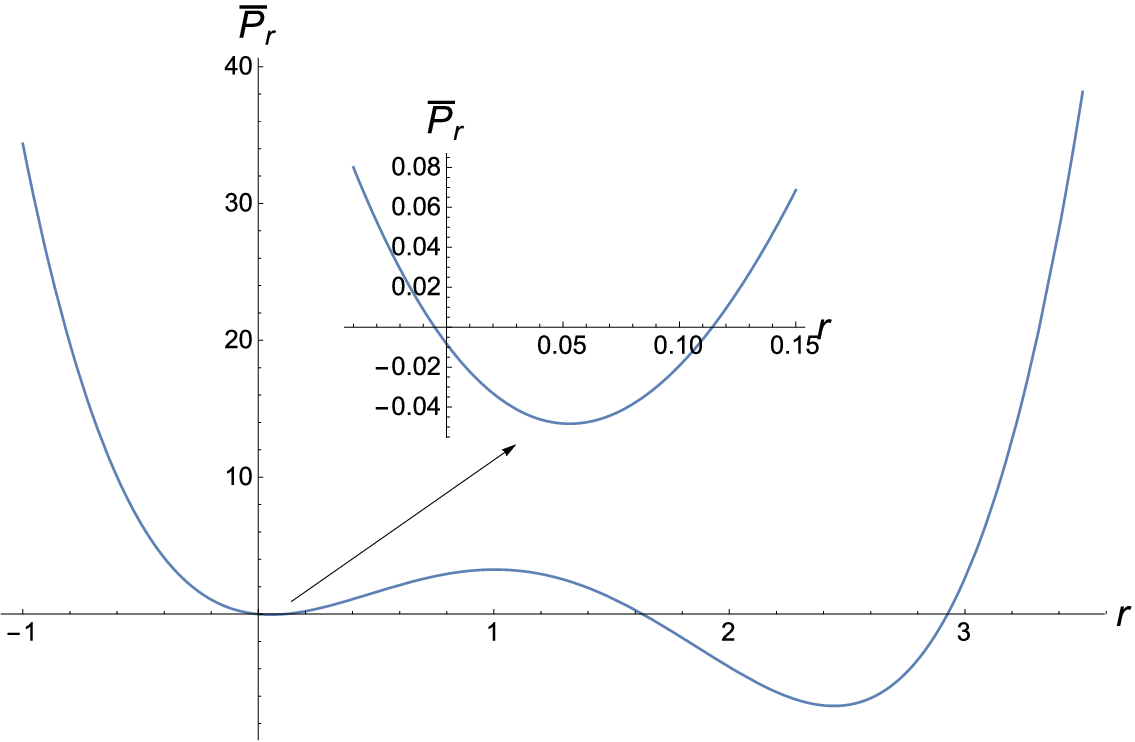}  } \\
\subfloat[]
     {\label{example_of_bound_orbit_energy_greater_than_one_3_2}
\includegraphics[width=0.45\linewidth]{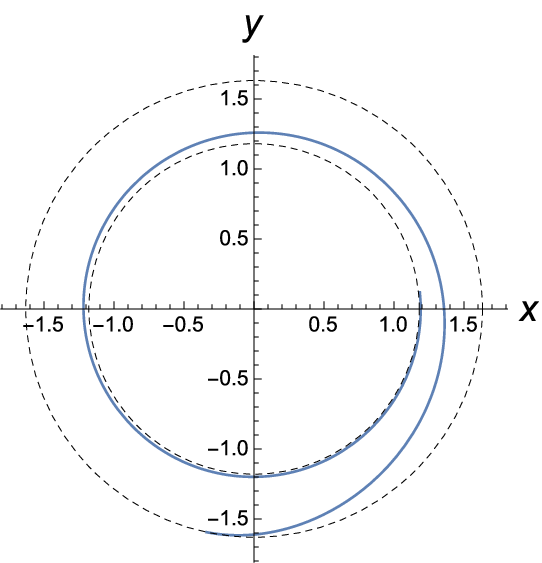}  }  \qquad
\subfloat[]
     {\label{example_of_escape_orbit_energy_greater_than_one_3_1}
\includegraphics[width=0.35\linewidth]{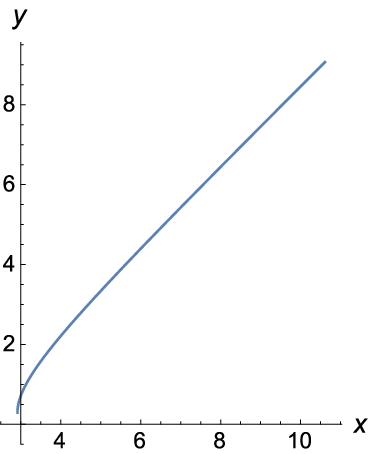}  }  \qquad
    \caption{Bound and escape orbits for $\bar{E}^2>1$ with parameters $M=1$, $a=0.9$, $Q=0.4$, $\ell=0.05$. Here, the roots of the radial potential $\bar{P}(r)$ satisfy $r_4<0<r_e^-<r_3<r_-<r_+<r_2<r_e^+<r_1$. Figure \ref{example_of_bound_orbit_energy_greater_than_one_3_2} illustrates the bound orbit in the ergoregion where $r_+<r <r_2<r_e^+$ for the test particle with $\bar{q}=-10$ and $\bar{E}=-2$. Here, the test particle cannot exit from ergoregion. The dashed circles indicate the bounds of the radial motion. Figure \ref{example_of_escape_orbit_energy_greater_than_one_3_1} illustrates the escape orbit in region where $r>r_1$ for the test particle with $\bar{q}=10$ and $\bar{E}=2$.} \label{example_of_bound_orbit_energy_greater_than_one_3}
\end{figure*}

\begin{figure*}[htp]
    \centering
 \subfloat[]
     {\label{example_of_bound_orbit_energy_less_than_one_1_1}
\includegraphics[width=0.6\linewidth]{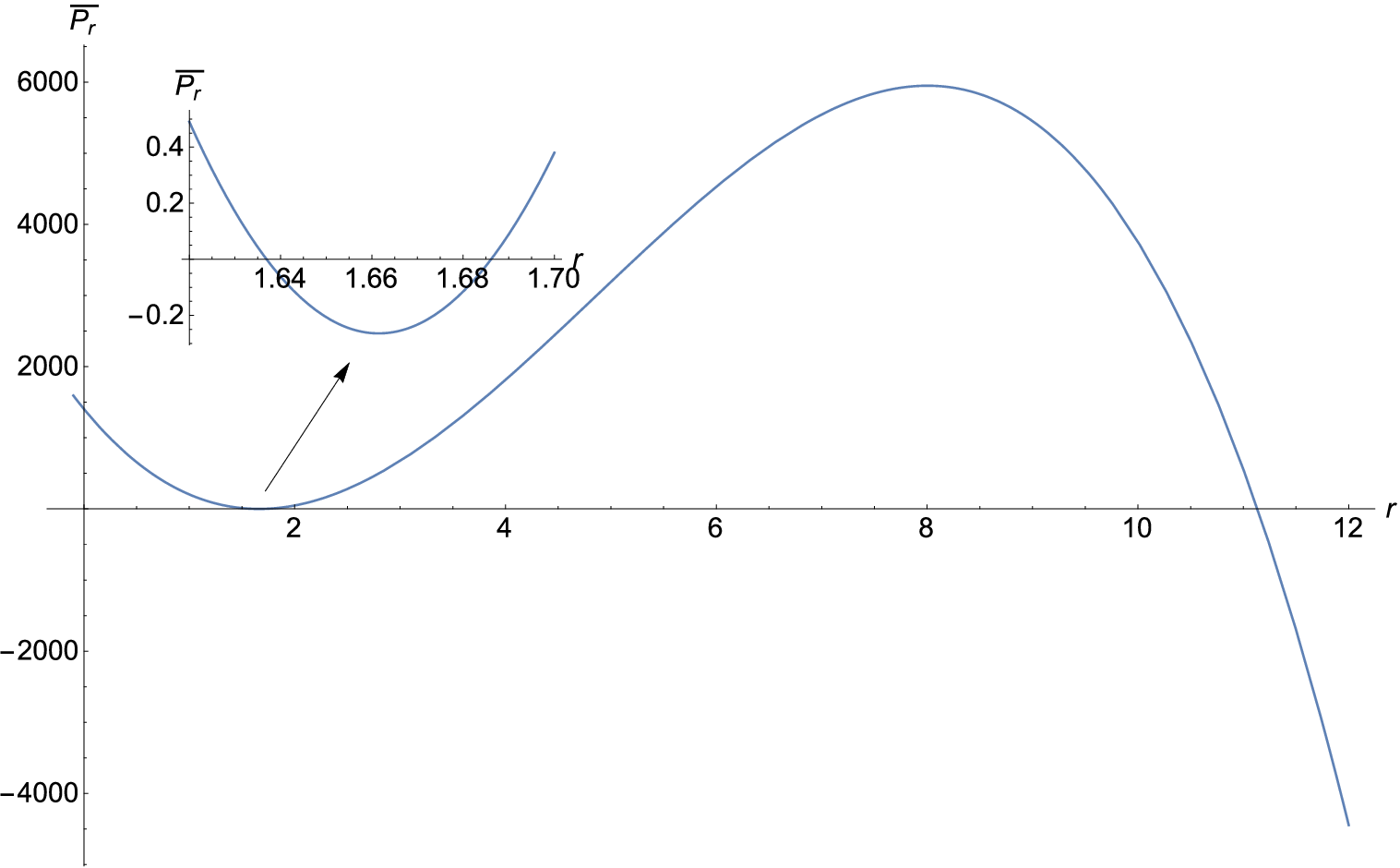}  }  \qquad
\subfloat[]
     {\label{example_of_bound_orbit_energy_less_than_one_1_2}
\includegraphics[width=0.3\linewidth]{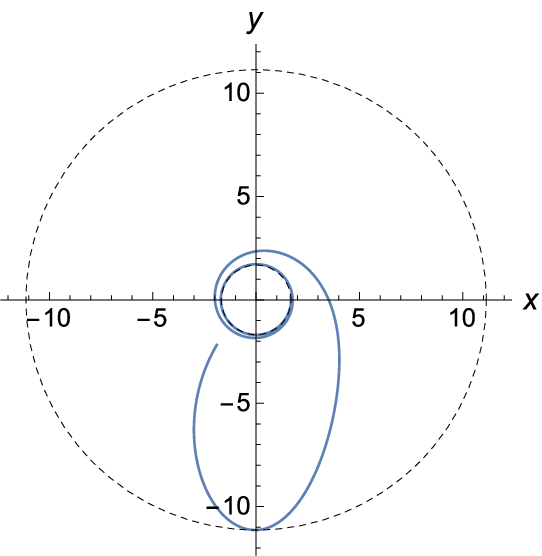}  }  \qquad
    \caption{Bound orbit for $\bar{E}^2<1$ with parameters $M=1$, $a=3.3$, $Q=5$ and $\ell=5.94$. Here, the roots of the radial potential $\bar{P}(r)$ satisfy $r_4<r_e^-<0<r_-<r_+<r_3<r_2<r_e^+<r_1$. Figure \ref{example_of_bound_orbit_energy_less_than_one_1_2} illustrates the bound orbit in the region where $r_2<r<r_1$ for the test particle with $\bar{q}=5$ and  $\bar{E}=0.9$  (direct orbits with $\bar{L}>0$). The test particle with positive energy enters into ergoregion and can exit from that region.The dashed circles indicate the bounds of the radial motion. On the $\bar{P}_r(r)$ graph the smallest root cannot be illustrated since $r_4=-240.773$ is out of the scale range of the radial coordinate.}\label{example_of_bound_orbit_energy_less_than_one_1}
\end{figure*}

\begin{figure*}[htp]
    \centering
\subfloat[]
     {\label{example_of_bound_orbit_energy_less_than_one_2_1}
\includegraphics[width=0.45\linewidth]{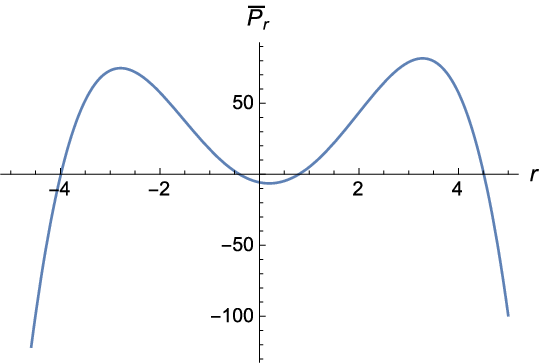}  }  \qquad    
\subfloat[]
     {\label{example_of_bound_orbit_energy_less_than_one_2_2}
\includegraphics[width=0.4\linewidth]{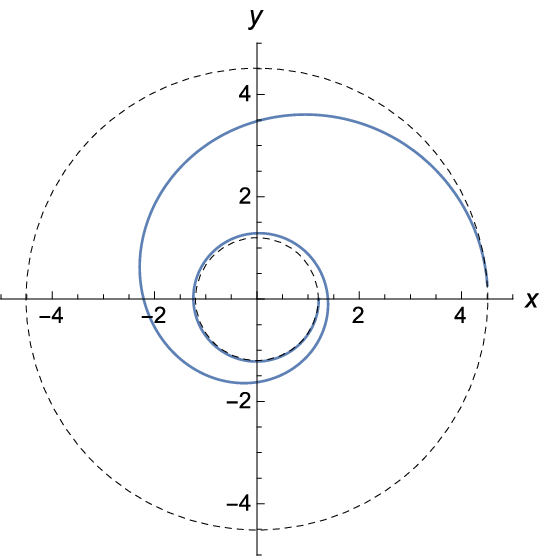}  }     
    \caption{Bound orbit for $\bar{E}^2<1$ with parameters $M=1$, $a=0.9$, $Q=0.4$ and $\ell=0.1$. Here, the roots of the radial potential $\bar{P}(r)$ satisfy $r_4<r_3<0<r_e^-<r_2<r_-<r_+<r_e^+<r_1$. Figure \ref{example_of_bound_orbit_energy_less_than_one_2_2} illustrates the bound orbit in the region where $r_+<r<r_1$ for the test particle with $\bar{q}=18.74$ and $\bar{E}=0.073$ (retrograde orbits with $\bar{L}<0$). Again the test particle with positive energy enters into ergoregion and can exit from that region. The dashed circles indicate the bounds of the radial motion.}\label{example_of_bound_orbit_energy_less_than_one_2}
\end{figure*}

\begin{figure*}[htp]
    \centering
    \subfloat[]
     {\label{example_of_bound_orbit_energy_equal_to_one_1_1}
\includegraphics[width=0.6\linewidth]{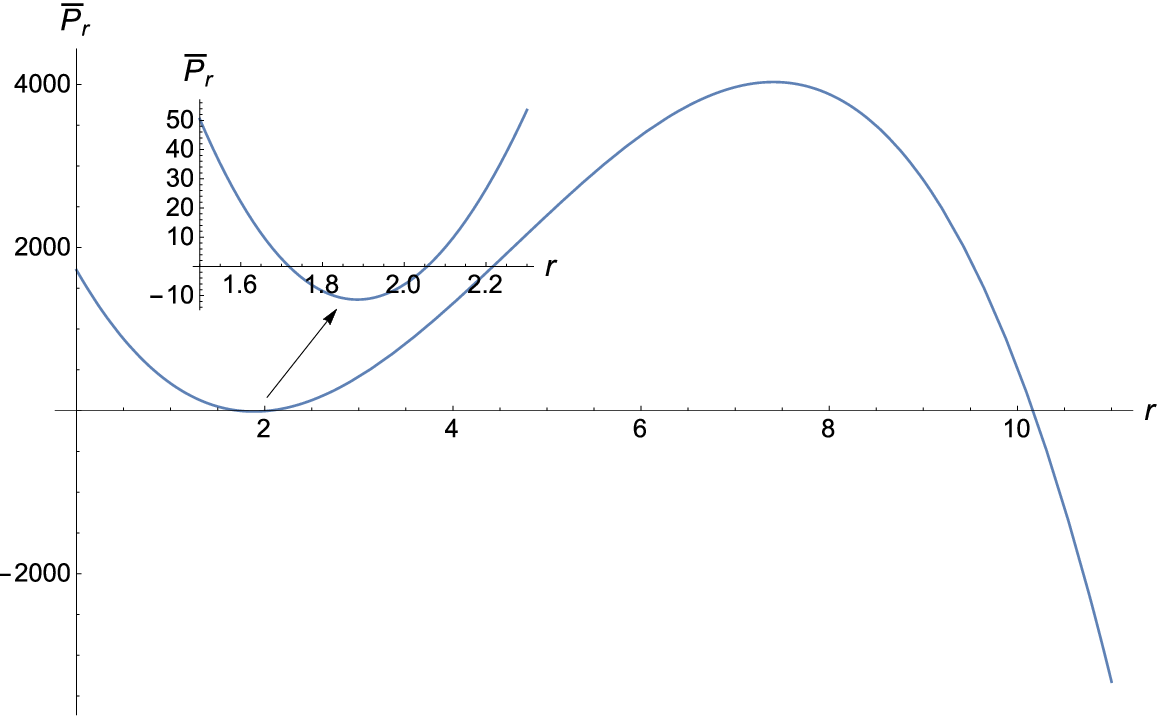}  }  \qquad
     \subfloat[]
     {\label{example_of_bound_orbit_energy_equal_to_one_1_2}
\includegraphics[width=0.3\linewidth]{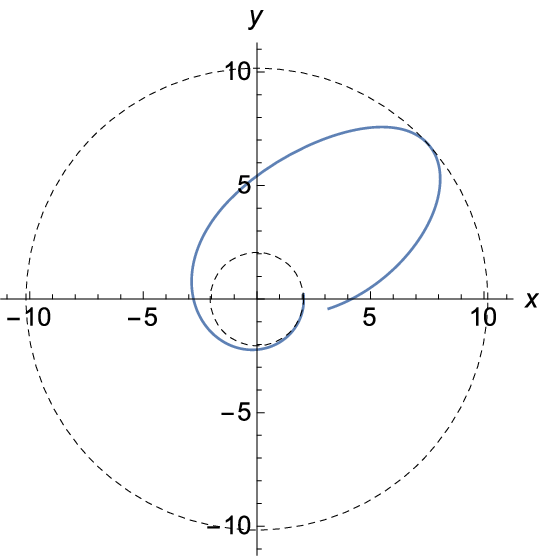}  } \\
 \subfloat[]
     {\label{example_of_escape_orbit_energy_equal_to_one_1}
\includegraphics[width=0.6\linewidth]{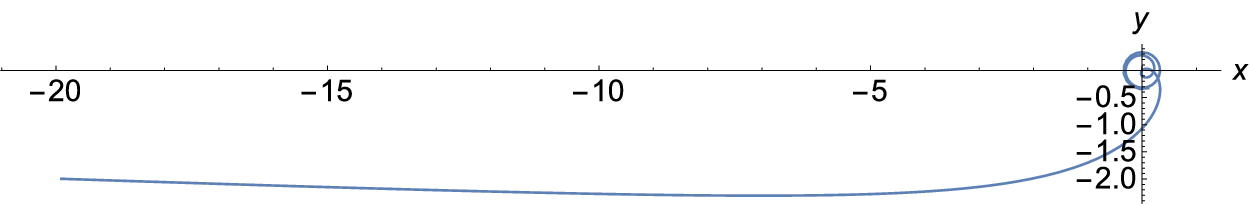}  } 
    \caption{Bound and escape orbits for $\bar{E}=1$ with parameters $M=1$, $a=3.4$, $Q=5$ and $\ell=6$. Here, the roots of the radial potential $\bar{P}(r)$ satisfy $r_e^-<0<r_-<r_+<r_3<r_2<r_e^+<r_1$. Figure \ref{example_of_bound_orbit_energy_equal_to_one_1_2} illustrates the bound orbit in the region where $r_2<r<r_1$ for the test particle with $\bar{q}=5$. For the bound motion, test particle with positive unity energy can enter into ergoregion where $r_+<r_e^+$ and can exit from that region. The dashed circles indicate the bounds of the radial motion. Figure \ref{example_of_escape_orbit_energy_equal_to_one_1} illustrates the crossover escape orbit in region where $-\infty<r<r_-$ for the test particle with $\bar{q}=5$. Here the test particle can enter into ergoregion where $r_e^-<r_- $ and can escape from that region to negative infinity.}\label{example_of_bound_orbit_energy_equal_to_one_1}
\end{figure*}

\noindent As a final remark, using analytical solutions presented above, we obtain trajectories of charged test particle for bound and escape orbits over the equatorial plane for $\bar{E}^2>1$, $\bar{E}^2<1$ and $\bar{E}^2=1$. These are illustrated in Figures \ref{example_of_bound_orbit_energy_greater_than_one_3}-\ref{example_of_bound_orbit_energy_equal_to_one_1}.

\subsection{Calculation of the perihelion shift for a bound orbit}

\noindent Here, to get an expression for the perihelion shift for a bound orbit, we consider that the motion in the radial direction is bounded in the interval $ r_2 \leq r \leq r_1$. Then, one can evaluate the fundamental period $\Lambda_r$ for the radial motion as
\begin{equation}
\Lambda_r=2 \int_{r_2}^{r_1} \frac{dr}{\sqrt{\bar{P}_r(r)}}=2\int_{y_0}^{\infty} \frac{dy}{\sqrt{\bar{P}_3(y)}} \label{observable_1}
\end{equation}
where $\bar{P}_3 (y)= 4 y^3 - g_2 y - g_3$ with $g_2$ and $g_3$ introduced in (\ref{analytic_13}). The integral can be calculated via the transformation
\begin{equation}
\xi=\frac{1}{\kappa} \left( \frac{\bar{y}_2-\bar{y}_3}{y-\bar{y}_3}\right)^{1/2} \label{observable_2}
\end{equation}
where $\bar{y}_1$, $\bar{y}_2$ and $\bar{y}_3$ correspond to the roots of the polynomial $\bar{P}_3 (y)= 0$ (ordered as $\bar{y}_3<\bar{y}_2<\bar{y}_1$) with $\kappa^2=\frac{\bar{y}_2-\bar{y}_3}{\bar{y}_1-\bar{y}_3}$. We also choose $y_0=\bar{y}_1$. Then one gets the radial period as
\begin{equation}
\Lambda_r=\frac{2}{\sqrt{\bar{y}_1-\bar{y}_3}} K(\kappa) \label{observable_3}
\end{equation}
where $K(\kappa)$ denotes the complete elliptic function with modulus $\kappa$. Then, one can also evaluate the corresponding angular frequency
\begin{equation}
\Upsilon_r=\frac{2 \pi}{\Lambda_r}=\frac{\pi \sqrt{\bar{y}_1 -\bar{y}_3}}{K(\kappa)} \label{observable_4}
\end{equation}
for the radial motion. Furthermore, one can obtain the angular frequencies $\Upsilon_{\varphi} $ and $\Upsilon_t $ for the $\varphi$-motion and $t$-motion respectively  from the solutions of $\varphi(\lambda)$ and $t(\lambda)$. By using the arguments exposed in \cite{drasco}, one can notice that the solutions $\varphi(\lambda)$ and $t(\lambda)$ can both be written in the forms
\begin{equation}
\varphi(\lambda)= \Upsilon_{\varphi} \left( \lambda - \lambda_0\right) + \bar{\varphi} (\lambda) \label{observable_5}
\end{equation}
and
\begin{equation}
t(\lambda)= \Upsilon_{t} \left( \lambda -\lambda_0\right) + \bar{t} (\lambda), \label{observable_6}
\end{equation}
where $\Upsilon_{\varphi} $ and $\Upsilon_t $ correspond to frequencies with respect to time parameter $\lambda$ for $\varphi$-motion and $t$-motion respectively. From these two solutions, one can get the corresponding angular frequencies as
\begin{eqnarray}
\Upsilon_{\varphi}&=& a \left[  \left( \bar{E} \ell^2 + \frac{1}{2 \bar{E}} \left( \ell^2- Q^2\right) \right) \omega_0 
+  \left( \frac{M}{\bar{E}}- \bar{q} Q \right) \bar{\omega}_0 + \left( \frac{2 \bar{E}^2 -1}{2\bar{E}}\right)  \tilde{\omega}_0
 \right] \\
&+&  a \sum_{i=1}^2  \sum_{j=1}^2   \left[  \left( \bar{E} \ell^2 + \frac{1}{2 \bar{E}} \left( \ell^2- Q^2\right) \right) \omega_i 
+  \left( \frac{M}{\bar{E}}- \bar{q} Q \right) \bar{\omega}_i+ \left( \frac{2 \bar{E}^2 -1}{2\bar{E}}\right)  \tilde{\omega}_i
 \right] \frac{\zeta (y_{ij}) }{ \wp^{\prime}  (y_{ij})} \nonumber 
\end{eqnarray}
and
\begin{eqnarray}
\Upsilon_{t}&=&\left[ \left( \bar{E} \ell^2 (a^2+\ell^2)+\frac{a^2 \ell^2}{\bar{E}}-\frac{a^2 Q^2}{2 \bar{E}}\right)\omega_0+\left( \frac{Ma^2}{\bar{E}}-(a^2+\ell^2)\bar{q}Q\right)\bar{\omega}_0\right.\nonumber\\
&&\left.+\bar{E}(a^2+2 \ell^2)\tilde{\omega}_0-\bar{q}Q\hat{\omega}_0+\bar{E}\check{\omega}_0\right]\nonumber\\
&&+\sum_{i=1}^3  \sum_{j=1}^2 \frac{(\bar{E}\check{\omega}_i-\bar{q}Q\hat{\omega}_i)}{ \wp^{\prime}  (y_{ij})}  \zeta (y_{ij})  -\bar{E} \check{\omega}_4 \sum_{j=1}^2 \frac{1}{\wp^{\prime \, 2}(y_{3j})} \left( \wp(y_{3j})+\frac{\wp^{\prime \prime}(y_{3j})}{\wp^{\prime}(y_{3j})}\right)\nonumber\\
&&+\sum_{i=1}^2  \sum_{j=1}^2 \frac{\zeta (y_{ij}) }{ \wp^{\prime}  (y_{ij})} \left[ \left( \bar{E}\ell^2(a^2+\ell^2)+\frac{a^2\ell^2}{\bar{E}}-\frac{a^2Q^2}{2\bar{E}}\right)\omega_i\right.\nonumber\\
&&\left.+\left( \frac{a^2 M}{\bar{E}}-(a^2+\ell^2)\bar{q}Q\right)\bar{\omega}_i+\bar{E}(a^2+2\ell^2)\tilde{\omega}_i\right].
\end{eqnarray}

\noindent Finally, as is also outlined in \cite{drasco}, \cite{fujita} and \cite{hackmann2}, the angular frequencies obtained using the time parameter $\lambda$ can be related to the angular frequencies $ \Omega_r$ and $\Omega_{\varphi}$ obtained with respect to a distant observer time as
\begin{equation}
\Omega_r=\frac{\Upsilon_{r}}{\Upsilon_{t}}, \qquad \Omega_{\varphi}=\frac{\Upsilon_{\varphi}}{\Upsilon_{t}}. \label{observable_13}
\end{equation}
 Obviously, these frequencies are not equal to each other. Therefore, it enables us to calculate the perihelion shift in the form
 \begin{equation}
\Omega_{perihelion}=\Omega_{\varphi}-\Omega_{r}. \label{observable_14}
\end{equation}

\noindent It is clear that, the perihelion shift explicitly depends on the NUT parameter and other physical spacetime parameters as well as the charge and the energy of the test particle. If one makes a comparison of this theoretical expression with those provided in astronomical observations, one can possibly comment about the existence of the NUT parameter in the real physical world \cite{bhattacharyya}. Although we couldn't provide a numerical value for the perihelion precision, one can see that the NUT parameter and the charge of the test particle have a definite influence on the perihelion shift.

\section{Conclusion}

In this study, we have comprehensively examined the equatorial orbits of a charged test particle in the background of KNTN spacetime. Having obtained the governing orbit equations, we have made an analysis of possible orbit types that would come out via the analysis of radial potential $\bar{P}_r(r)$. We have accomplished a comprehensive investigation of equatorial orbit types with respect to the value of the energy $\bar{E}$ of the test particle and the form of the radial potential $\bar{P}_r(r)$. To see the effect of NUT parameter on the formation of possible  orbit configurations, we have made a graphical analysis with respect to change in NUT parameter. Next, by using Descartes' rule of sign, we have made a detailed investigation of the existence of bound orbits in the causality-preserving region (outside the event horizon where $r>r_+$) and obtained required conditions for the existence (or non-existence) of them. In addition, we have investigated the required conditions for the existence of equatorial circular orbits for charged and uncharged particles. 
It is seen that the relations (\ref{circular_5}) and (\ref{circular_6}) determine the existence of equatorial circular orbits for a charged test particle in KNTN spacetime while the expressions (\ref{circular_9}) and (\ref{circular_10}) fix the conditions for the existence of circular orbits for an uncharged test particle.  

\noindent As a further remark, we have worked out elliptical Newtonian orbits over equatorial plane in presence of NUT charge. We have explicitly seen that unlike the other spacetime parameters ($a$, $Q$ and $M$), the effect of the NUT parameter can be observed as the general relativistic correction (improvement) to classical Newtonian orbits. Finally, to obtain trajectory of charged test particle in equatorial KNTN spacetime, we have solved orbit equations and obtained the exact analytical solutions in terms of Weierstrass $\wp$, $\sigma$ and $\zeta$ functions. Using these analytical solutions, we have also plotted trajectories of charged test particle for bound and escape orbits in the regions where $r<r_-$ and $r>r_+$. In addition, as a physical observable, we have calculated the perihelion shift for a bound orbit over the equatorial plane where it obviously depends on the NUT and other physical parameters. We believe that, one can surely comment on the existence of the NUT parameter in the universe if the theoretical expression for the perihelion shift is compared with the numerical values provided through astronomical observations. Also we can comment that a comprehensive investigation of gravitational waves can lead to detection of NUT charge in the universe as well.

\noindent For a future study, it would also be physically interesting to investigate the equatorial orbits of the charged test particles in rotating Taub-NUT spacetimes with cosmological constant (Kerr-Newman-Taub-NUT-(A)dS spacetimes). In particular, the investigation of the existence of equatorial circular orbits in such a spacetime deserves further study to see the effect of the cosmological constant. These are devoted to future research.

\begin{acknowledgements}
We would like to thank anonymous reviewers for their suggestions and comments. 
\end{acknowledgements}

\end{document}